\documentclass{ametsocV6.1}

\title{Machine learning models for daily rainfall forecasting in Northern Tropical Africa using tropical wave predictors}

\authors{Athul Rasheeda Satheesh, Peter Knippertz and Andreas H. Fink\correspondingauthor{Athul Rasheeda Satheesh, athul.satheesh@kit.edu}}

\affiliation{{Institute of Meteorology and Climate Research Troposphere Research, Karlsruhe Institute of Technology, Karlsruhe, Germany}}

\abstract{Numerical weather prediction (NWP) models often underperform compared to simpler climatology-based precipitation forecasts in northern tropical Africa, even after statistical postprocessing. AI-based forecasting models show promise but have avoided precipitation due to its complexity. Synoptic-scale forcings like African easterly waves and other tropical waves (TWs) are important for predictability in tropical Africa, yet their value for predicting daily rainfall remains unexplored. This study uses two machine-learning models—gamma regression and a convolutional neural network (CNN)—trained on TW predictors from satellite-based GPM IMERG data to predict daily rainfall during the July-September monsoon season. Predictor variables are derived from the local amplitude and phase information of seven TW from the target and up-and-downstream neighboring grids at 1-degree spatial resolution, which then undergo selection via a gradient-boosting regression algorithm. The ML models are combined with Easy Uncertainty Quantification (EasyUQ) to generate calibrated probabilistic forecasts and are compared with three benchmarks: Extended Probabilistic Climatology (EPC15), ECMWF operational ensemble forecast (ENS), and a probabilistic forecast from the ENS control member using EasyUQ (CTRL EasyUQ). The study finds that downstream predictor variables offer the highest predictability, with downstream tropical depression (TD)-type wave-based predictors being most important. Other waves like mixed-Rossby gravity (MRG), Kelvin, and inertio-gravity waves also contribute significantly but show regional preferences. ENS forecasts exhibit poor skill due to miscalibration. CTRL EasyUQ shows improvement over ENS but only marginal enhancement over EPC15 in land regions. Both gamma regression and CNN forecasts significantly outperform benchmarks in tropical Africa.}

\begin{document}

\maketitle

%
%
%
%
%
%

%

\section{Introduction \label{intro}}

While the concept of numerical weather prediction (NWP) dates back to pre-World War II times, it was not until the late 1950s that a fully functional NWP model was successfully deployed \citep{shuman1989history, bauer2015quiet}. A main challenge encountered during the initial phase was the scarcity of observational data for initializing these models. With the advent of the satellite era in the late 1970s, the issue of sparse observations gradually eased. Today, advancements in technology have propelled major operational weather services worldwide to routinely issue weather forecasts across diverse spatial and temporal resolutions at varying lead times. 

Despite these advances, tropical precipitation forecasts, notably for tropical Africa, have yet to achieve the desired level of accuracy, already achieved for the midlatitudes \citep[e.g.,][]{haiden2012intercomparison, vogel2018skill, vogel2020skill}. The inherent reduced predictability, in particular of meso-scale weather in tropical regions, characterized by faster error growth compared to midlatitudes \citep{judt2020atmospheric}, presents formidable challenges, which is exacerbated by limited in-situ observations in tropical Africa \citep{parker2008amma}.  Contemporary NWP ensemble precipitation forecasts struggle to surpass the predictive accuracy of simpler climatology-based forecasts, even after statistical postprocessing \citep{vogel2018skill, walz2021imerg}. While NWP forecasts demonstrate some skill in predicting orographically triggered precipitation \citep{vogel2020skill}, such capability is of limited utility in areas like tropical West Africa, where the landscape is predominantly flat.

\begin{figure*}[!htp]
    \centering
    \noindent\includegraphics[width=\textwidth]{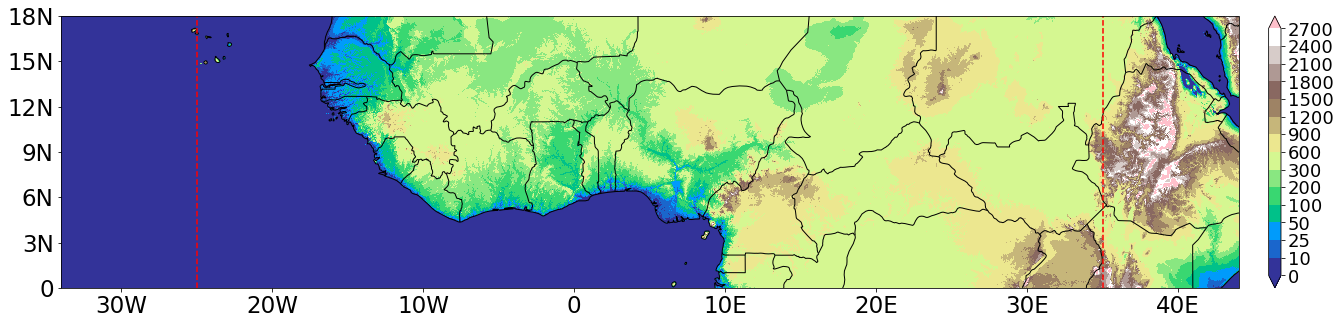}
    \caption{Geographical overview of the analysis domain. Shading indicates altitude in metres. The red dashed lines demarcate the nested core domain ($25^\circ \text{W}-35^\circ \text{E}$ in longitude and $0^\circ -18^\circ \text{N}$ in latitude) where the machine learning forecasts are issued. Modified from \citet{rasheeda2023sources}.}
    \label{fig:domain}
\end{figure*}

Additionally, forecasting convection initiation is challenging due to the inability of current NWP models to resolve many relevant small-scale processes. Hence, NWP models parameterize convection in a simplified form, another potential factor for reduced skill. In tropical Africa, especially in the Sahel, organized mesoscale convective systems \citep[MCSs;][]{mathon2002mesoscale} play a crucial role in annual rainfall, which is often inadequately represented in convection parameterization \citep{marsham2011importance, pante2019resolving}. Convection-permitting (CP) models partially address this issue, but accurately predicting the location and intensity of rainfall remains a challenge, despite the high computational expenses invested in them \citep{cafaro2021convection, senior2021convection}. These limitations in forecast performance pose serious challenges, for agriculture — an essential economic sector in tropical Africa, vital services such as flood and drought warnings, and infrastructure management like dams for hydropower generation.

A potential avenue for addressing these challenges are data-driven, i.e., statistical models. Early attempts using Markov models predate the widespread adoption of NWP models \citep{gabriel1962markov,gates1976markov}. While they faced lower popularity initially due to their non-reliance on explicit atmospheric dynamics, a gradual acceptance within the meteorological community, aided by advancements in statistical computation and recent progress in machine learning (ML) and artificial intelligence (AI), has been observed. Models like FourCastNet, GraphCast and Pangu-Weather \citep{pathak2022fourcastnet, lam2023learning, bi2023accurate} showcase the potential of AI in meteorology, offering comparable performance to NWP models with cost and time advantages. Nevertheless, accurate rainfall prediction remains challenging due to its sparse and non-Gaussian nature \citep{lam2023learning}.

Studies by \citet{vogel2021statistical} and \citet{rasheeda2023sources} have highlighted that a simple logistic regression model can effectively predict daily rainfall occurrences during the summer monsoon season in tropical Africa. These models, trained on past rainfall data, have shown superior performance compared to climatology and NWP-based forecasts. Their success lies in identifying predictor grid points strongly correlated with the target grid point in the preceding days. Building upon this concept, \citet{walz2024physics} introduced a convolutional neural network (CNN) model trained on gridded rainfall data, combined with Easy Uncertainty Quantification \citep[EasyUQ;][]{walz2024easy} to generate probabilistic forecasts of daily rainfall amounts. Their approach outperforms ensemble precipitation forecasts from the ECMWF and climatology-based forecasts across tropical Africa. Additionally, \citet{antonio2024postprocessing} utilized a conditional generative adversarial network (cGAN) with quantile mapping to produce skilful precipitation forecasts over East Africa. These successes highlight the potential of deep-learning models in capturing complex relationships between precipitation and dynamical features.

A potential predictor for statistical models are tropical waves (TWs), which play a crucial role in governing coherent rainfall patterns in tropical regions \citep{matsuno1966quasi, kiladis2009convectively, knippertz2022intricacies}. Numerous investigations have been undertaken to understand the intricate relationship between synoptic-mesoscale rainfall dynamics and TWs across tropical Africa \citep[e.g.,][]{sinclaire2015synoptic, mekonnen2016mechanisms, ayesiga2021observed}. \citet{schlueter2019systematic_b, schlueter2019systematic_a} conducted systematic comparisons of the impacts of various TWs, encompassing convectively coupled equatorial waves \citep[CCEWs;][]{wheeler1999convectively}, the Madden-Julian Oscillation \citep[MJO;][]{madden1971detection, madden1972description}, and tropical depression (TD)\footnote{TDs serve as proxies for African easterly waves in northern tropical Africa}-like waves.  

These investigations have revealed that fast-propagating waves, such as Kelvin waves and AEWs \citep{fink2003spatiotemporal}, influence precipitation variability at daily to sub-daily timescales, whereas slower waves, such as the MJO and Equatorial Rossby (ER) waves, modulate rainfall patterns over 7 to 20 days. However in \citet{schlueter2019systematic_b, schlueter2019systematic_a}, the authors also argue that these low-frequency wave modes, during their wet (dry) phase, tend to amplify (suppress) the activity of faster wave modes, thereby indirectly impacting daily rainfall. Furthermore, \citet{lawton2022influence} document the potential of Kelvin waves to trigger AEWs in the Sahel region, which further initiate and modulate MCSs. As shown in \citet{schlueter2019systematic_a}, TWs coherently modify the thermodynamical environment for deep convection, potentially improving predictability. \citet{vogel2021statistical} and \citet{rasheeda2023sources} attribute the coherently propagating rainfall correlation fields identified in their studies to propagating features like AEWs in the Sahel and westward-propagating cyclonic vortices over the Gulf of Guinea \citep{knippertz2017meteorological}. 

Despite these insights, operational rainfall forecasts have yet to directly utilize TWs as predictors. Hence, through training ML models exclusively on convectively coupled TWs\footnote{Following \citet{schlueter2019systematic_b}, we use the term TWs to refer to CCEWs, MJO and TD}, derived from the satellite-based gridded Global Precipitation Measurement Integrated MultisatellitE Retrievals (hereafter, GPM IMERG) precipitation product \citep{huffman2020integrated} to predict daily rainfall amounts in tropical Africa, we seek to bring this predictive potential to fruition.

To this end, the remainder of this paper is organized as follows: Section~\ref{data_methods} outlines the specifics of the observational and NWP-based forecast data used in this study. Subsequently, it provides a detailed description of the data preprocessing, development and selection of predictor variables, evaluation metrics and tools, and benchmarking strategies. Section~\ref{ml_models} explains the theory and configuration of the ML models employed for generating precipitation forecasts. Subsequently, in Section~\ref{results}, we first present the outcome of the predictor selection to identify suitable predictor variables across tropical Africa. We then present the forecast results, beginning with the examination of an illustrative grid point near Niamey ($13^\circ$N, $2^\circ$E), followed by an analysis spanning entire northern tropical Africa. Lastly, in Section~\ref{conclusions}, we offer a summary, discuss the main findings of this study, and provide a brief outlook.

\section{Data and Methods\label{data_methods}}
\subsection{Employed Data \label{data:employed_data}}
We utilize GPM IMERG V07 dataset (2001--2022) to represent precipitation, potentially benefiting from enhancements in detection, systematic bias correction, and reduction of random biases compared to its predecessor, V06 \citep{huffman2023imerg}. To ease the computational burden and minimize small-scale noise, we conservatively remap this dataset to a spatial resolution of $1^\circ \times 1^\circ$. The daily accumulated precipitation (from $06\text{UTC}$ to $06\text{UTC}$) is selected as the target variable, which is confined within the latitude band of $0^\circ$ to $18^\circ \text{N}$ and the longitudinal range of $25^\circ \text{W}$ to $35^\circ \text{E}$ (Fig.~\ref{fig:domain}), following \citet{rasheeda2023sources} and \citet{walz2024physics}. 

The TW-filtered precipitation, utilized in computing predictor variables, is derived from 6-hourly mean precipitation data spanning from $00\text{UTC}$ to $06\text{UTC}$. However, these are computed over a slightly broader zonal extent ($35^\circ \text{W}$ to $45^\circ \text{E}$) to facilitate the extraction of predictors from neighbouring grid points, as elaborated in Section \ref{data_methods}\ref{grad_boost}.

The forecasts from the ML models are compared with the operational 24-hr ensemble precipitation forecast (ENS) downloaded from ECMWF's Meteorlogical Archival and Retreival System (MARS; \url{https://www.ecmwf.int/en/forecasts/access-forecasts/access-archive-datasets}). ENS comprises 50 perturbed forecasts and one control forecast at a spatial resolution of $0.25^\circ$ $\times$ $0.25^\circ$. We conservatively remap this to the spatial resolution of $1^\circ$ $\times$ $1^\circ$ and compute daily accumulation to match the grid spacing of the GPM IMERG rainfall product.

\begin{table}[!htp]
\caption{The wavenumber, period and equivalent depths used to filter tropical depression (TD), mixed Rossby-gravity (MRG), Madden-Julian oscillation (MJO), Kelvin, $\text{n}=1$ (westward) inertio-gravity (IG1), $\text{n}=1$ equatorial Rossby (ER), and $\text{n}=0$ eastward inertio-gravity (EIG) wave. Based on \citet{janiga2018subseasonal}.}\label{tab:tw_filter}
\begin{center}
\begin{tabular}{cccc}
\hline\hline
Wave   & Wavenumber & Period (days) & Equivalent depth (m)\\
\hline
TD     & $-20\text{ to }-6$   & $2.5\text{ to }5$               & $-$    \\
MRG    & $-10\text{ to }-1$   & $3\text{ to }8$                 & $8\text{ to }90$ \\
MJO    & $0\text{ to }9$      & $30\text{ to }60$               & $-$    \\
Kelvin & $1\text{ to }20$     & $2.5\text{ to }20$              & $8\text{ to }90$ \\
IG1    & $-20\text{ to }-1$   & $1.4\text{ to }2.5$             & $8\text{ to }90$ \\
ER     & $-10\text{ to }-1$   & $9\text{ to }72$                & $8\text{ to }90$ \\
EIG    & $0\text{ to }14$     & $1.82\text{ to }5$              & $8\text{ to }90$ \\
\hline
\end{tabular}
\end{center}
\end{table}

\subsection{Predictor pre-processing \label{pre_process}}
To input the data into the proposed ML models, it is necessary to first pre-process them in the following way: Seven TWs (see Table~\ref{tab:tw_filter}) are extracted from the 6-hourly precipitation data using an FFT-based wave-filter algorithm \citep[WK99;][]{wheeler1999convectively} and select 00UTC-06UTC times to compute predictor variables. 

Then, the predictor variables are computed from the filtered wave modes by following these steps: 1) The wave-filtered precipitation is divided into 'training-testing' pairs in a `leave-one-out' cross-validation mode before filtering, as in \citet{vogel2021statistical} and \citet{rasheeda2023sources}. 2) To reduce edge effects from FFT transforms, the training dataset has its first and last three years removed after filtering. Similarly, three years of zeroes are padded on either side of every testing set before filtering, which are not considered post-filtering. 3) As explained in the following subsection, the local phases and amplitudes at every grid point are computed from all training-testing pairs. To forecast each year between 2007 and 2019, we utilize data from 2004 to 2019, excluding the specific year to be forecast.  This method ensures robust predictions by utilizing a comprehensive dataset spanning 15 years and the outputs are not be affected by the specific nature (whether overly wet or dry) of the testing year.

\subsection{Local phase-amplitude computation \label{local_phase_amp}} 
Since TWs are synoptic- to planetary-scale features, using them as predictors at individual grid points is sub-optimal. Therefore, we calculate the local phases and amplitudes of each wave mode at every grid point using a modified method from \citet{yasunaga2012differences} and \citet{van2016modulation}. Summarizing: 1) Compute and normalize the filtered wave signal and its time derivative at any location. 2) Plot a phase-amplitude diagram (Fig.~\ref{fig:phase_amplitude}), with the filtered wave signal on the horizontal axis and its time derivative on the vertical axis. Each scatter point represents a timestep. 3) Define the radial distance of a scatter point as the local amplitude $A$ and its angle with the positive x-axis as the local phase ($\theta$). 4) Calculate the predictor variable, 'phase-adjusted wave amplitude' (PWA), for each TW type at every timestep `$t$' by combining the amplitude and $\theta$ as:

\begin{equation} \label{eq
}
\text{PWA(t)\textsubscript{wave}} = \text{A(t)\textsubscript{wave}}\times \cos{\theta(\text{t})\textsubscript{wave}}
\end{equation}

This method incorporates both wave amplitude and phase into the predictors while reducing the number of predictors needed to train ML models by half\footnote{The \texttt{Python} software package for computing local phase-amplitude is available at \url{https://github.com/athulrs177/phase_amplitude.git}}.

\begin{figure}[!ht]
    \centering
    \noindent\includegraphics[width=0.48\textwidth]{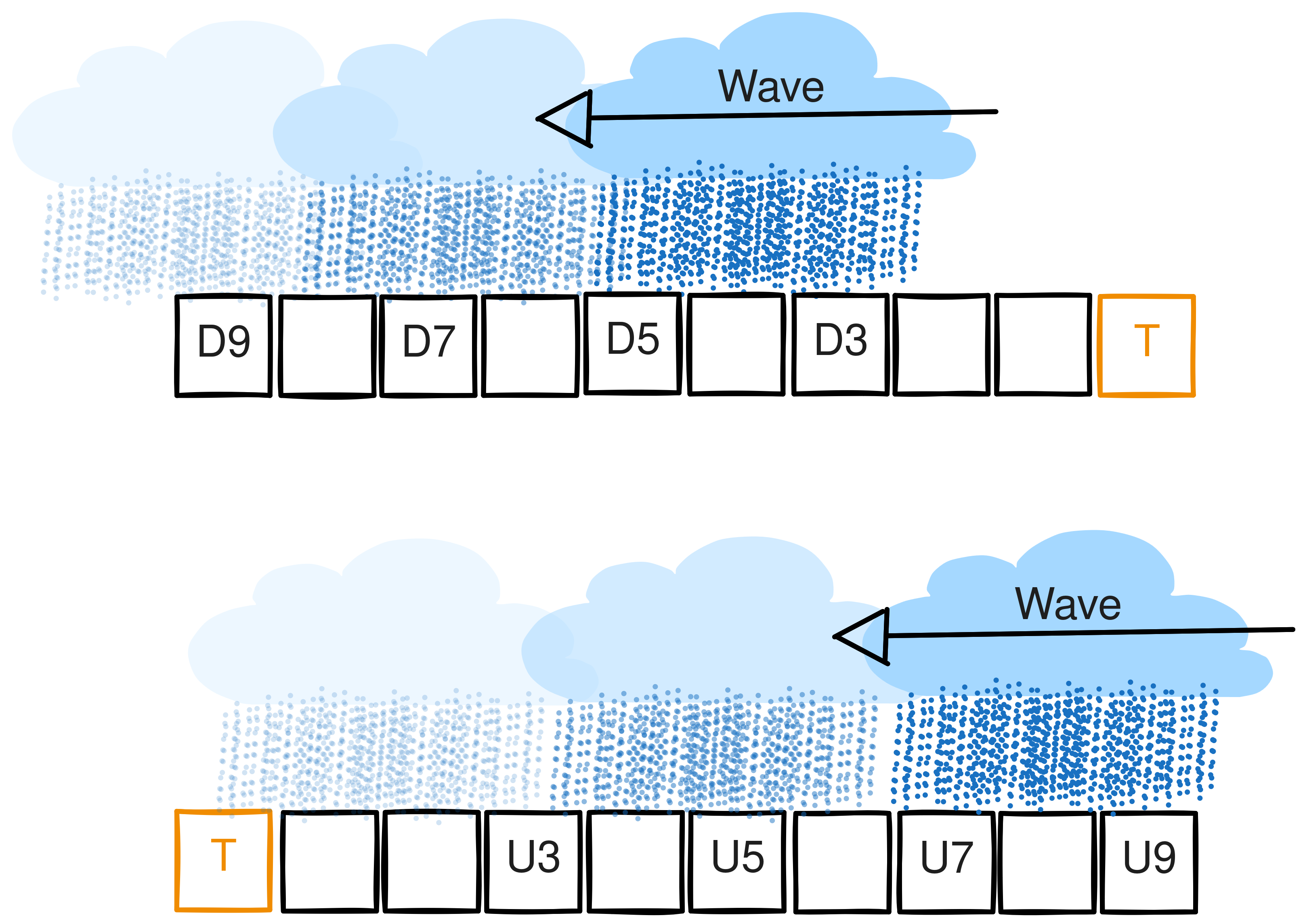}
    \caption{Schematic illustration of the mechanism behind the predictor selection algorithm. \textit{DN} (\textit{UN}) refers to \textit{N} grid points downstream (upstream) of the target grid point (\textit{T}) in the direction of the considered wave's propagation. The top (bottom) sketch shows how downstream (upstream) predictors offer predictability (see Section \ref{data_methods}\ref{grad_boost} for details). Sketches in lower opacity indicate the future locations of the propagating rainfall system modulated by a TW.}
    \label{fig:upstream_downstream_schematic}
\end{figure}

\subsection{Predictor selection via gradient-boosting regression \label{grad_boost}}
Following the methodology employed by \citet{rasheeda2023sources} and \citet{vogel2021statistical}, we have designed the ML models to issue separate forecasts at each grid point. To train models at a specific target grid point, potential predictor variables (i.e. the PWAs defined above) are identified from a total of nine grid points: four downstream grid points, the target grid point itself, and four upstream grid points (see Fig.~\ref{fig:upstream_downstream_schematic} for details). The downstream and upstream grid points are calculated based on the zonal propagation direction of the considered TW type. As a result, they differ for wave modes propagating westward and eastward. Hence, for training a forecast model at any target grid point, we have a total of 63 potential predictors (7 wave modes $\times$ 9 predictors each).

A recent study by \citep{rasheeda2024statistical} demonstrated that using predictors from the target grid point and its closest neighboring points yielded good results. Including up to the ninth neighboring grid point (approximately 1000 km away) improved regional results but not OVER? the entire domain (not shown). Performance peaked at the ninth grid point, likely due to its distance being comparable to the half-wavelengths of the most active TWs in Africa (e.g., TDs). Incorporating all nine up-and-downstream and the target grid points improved overall domain performance but increased computational costs significantly. Using the third, fifth, seventh, and ninth closest upstream and downstream grid points, along with the target grid point, retained most predictive information while managing computational costs.

As understanding the relationship between predictor variables is crucial for interpreting statistical model results \citep{walz2024physics}, we conducted a multi-collinearity test on all 63 predictors at an exemplary grid point near Niamey ($13^\circ$N, $2^\circ$E; refer to Fig.~\ref{fig:feature_correlation}). The results of the test showed that high positive correlations only exist between the closest neighbouring grid points for a given wave mode, as these waves are expected to impact adjacent grid points similarly as they propagate. Correlations between different wave modes are negligible, except for the correlations of EIG with IG1 and Kelvin waves (approximately 0.25).

\begin{figure}[!ht]
    \centering
    \noindent\includegraphics[width=0.8\textwidth]{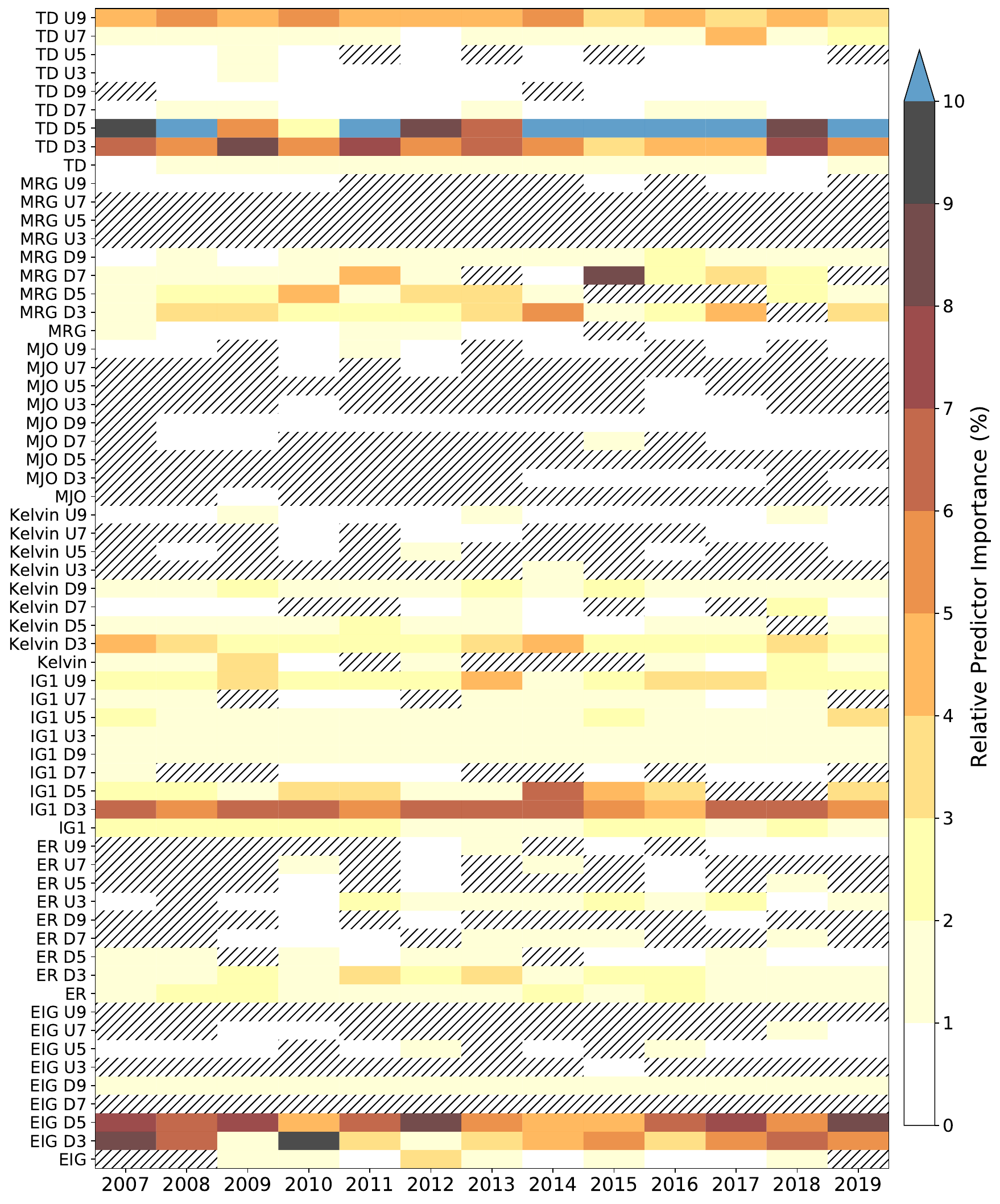}
    \caption{Relative importance of predictor variables at the grid point nearest to Niamey ($13^\circ$N, $2^\circ$E) for every year from 2007-2019. `Wave UN' (`Wave DN') refers to the wave predictor N grid points upstream (downstream) of the target in the direction of wave propagation. Just 'Wave' refers to the wave predictor at the target grid point. Shading indicates the relative predictor importance in percentage. `UN' and `DN' notations will be used throughout the text to refer to upstream and downstream predictors. Note that the sum of relative predictor importance from all 63 predictors is 100$\%$ every year. Only unhatched predictors are selected for training the forecast models.}
    \label{fig:feature_importance}
\end{figure}

When dealing with a large pool of potential predictors, training forecast models on all of them is often unnecessary. Therefore, a predictor-selection strategy is essential. Manually selecting suitable predictors for all grid points can be laborious and error-prone, and contradicts the motivation for using ML strategies. Thus, we present a semi-automated predictor selection algorithm using gradient-boosting regression\footnote{Implemented using the \texttt{eXtreme Gradient Boosting (XGBoost) version 1.7.5} \texttt{Python} package} \citep{friedman2001greedy, chen2016xgboost} to identify the most appropriate predictors for each target grid point. Summarizing: 1) Split the training dataset into two parts, and use the final year for validation. 2) Fit the gradient-boosting model and identify predictors that minimize the validation loss, assuming daily rainfall follows a gamma distribution \citep[e.g., ][]{coe1982fitting, wilks1990maximum, sloughter2007probabilistic, scheuerer2015statistical}. Thus, the \texttt{reg:gamma} option is used to estimate the loss function. 3) Train forecast models using only predictors that exceed a threshold of `\texttt{$0.80 \times \text{median}$}' of the relative predictor importance\footnote{Defined as \texttt{gain} in \texttt{XGBoost}} values. 4) Repeat this process for all training-testing splits between 2007 and 2019 (leave-one-out cross-validation). Fig.~\ref{fig:feature_importance} demonstrates the results by showing the relative predictor importance at the grid point near Niamey ($13^\circ \text{N}$, $2^\circ \text{E}$). We will now explain the outcome of the predictor selection algorithm using this example.

The activity of TD waves located 5 gridpoints (approximately 540 km) downstream of the Niamey grid point (TD D5) is consistently identified as the most significant predictor, except for 2009, 2010 and 2012 (see Fig.~\ref{fig:feature_importance}). It contributes sometimes more than 10$\%$ of the predictability of daily rainfall (blue boxes in Fig.~\ref{fig:feature_importance}). TD wave activity at D3 is another crucial predictor, contributing 5-9$\%$ of the predictability in most years. Kelvin D3 and D5, IG1 D3 and D5, ER D3 and D5, and EIG D3 and D5 are also consistently selected predictors in most years. Predictors from D3 and D5, collectively account for up to 50$\%$ of predictability every year. Figure~\ref{fig:feature_importance} also demonstrates that, in comparison, predictors from upstream grid points are generally less important. Notable exceptions are the TD U9 and IG1 U9 predictors, located 9 grid points (almost 1000 km) upstream of the target grid point, contributing about 6$\%$ and 3$\%$ of predictability, respectively, on average. It is also worth noting that all predictors based on slow large-scale waves, such as the MJO and the ER, are either not selected at all or contribute less than 1$\%$ and 2$\%$, respectively, per year on average.

Tropical Africa is the only region in the world where there is no dominant rainfall persistence. This is indicated by negative autocorrelation of one-day lagged daily precipitation anomalies \citep{roehrig2013present, spat2024autocorrelation}. Our analysis shows that downstream and some upstream predictors have a greater impact than the predictors at the target grid point. To expand on this further, we propose the following mechanism: Given that precipitation supported by the wet phase of a TW was observed at downstream grid points D3 and D5 at 00UTC, the target grid point will increasingly come under the influence of rainfall suppressing effects of the dry part of the TW during our forecast target period 06--06 UTC. Consequently, the strongest statistical relationship that we can employ for forecasting appears to be that of rain suppression upstream of a TW wet phase propagating away from the target grid point. On the other hand, rainfall modulated by a TW at upstream grid point U9 at 00 UTC will likely propagate towards the target grid point. As a result, PWA at upstream grid points can provide valuable information about incoming precipitation systems within the next 30 hours (6 hours lead time + 24 hours accumulation time). These cases are illustrated by a schematic in Fig.~\ref{fig:upstream_downstream_schematic}. 

\subsection{Evaluation metrics \label{metrics}}
\subsubsection{Taylor score \label{taylor_score}}
According to \citet{taylor2001summarizing}, the skill of a deterministic forecast can be evaluated based on two factors: how closely the forecast values match the observed values (low root mean squared error, RMSE) and how accurately the forecast represents observed patterns and trends (high correlation). The Taylor Score $S$ combines these two aspects and ranges from zero (least skilful) to one (most skilful):
\begin{equation} \label{eq:taylor_score}
    S=\frac{4(1+\rho)}{(\hat{\sigma} + \frac{1}{\hat{\sigma}})^2(1 + \rho_0)}
\end{equation}
where $\hat{\sigma}$ is defined as the forecast variance normalized with the observed variance, and $\rho_0$ is the maximum correlation possible assumed here to be 1. So as $\hat{\sigma}\to 1$ and $\rho\to 1$, $S\to 1$. On the other hand, when $\rho\to -1$ or $\hat{\sigma}\to \infty$, $S\to 0$. Interestingly, $S\to 0.5$, in the case of perfect variance and zero correlation.

\subsubsection{Continuous ranked probability (skill) score \label{crps}}
When the probabilistic prediction is presented as a cumulative distribution function, and the observations are scalar ($y_i$), such as in the case of the ensemble forecasts of rainfall accumulation, the continuous ranked probability score \citep[CRPS;][]{gneiting2007strictly} is the most widely used metric. The instantaneous CRPS is determined by computing the quadratic difference between the predicted cumulative distribution function ($F_i$) and the observed empirical cumulative distribution function \citep[$\mathbf{1}(x\geq y_i)$;][]{zamo2018estimation}. However, practically the CRPS is often used as a mean score defined as:

\begin{equation} \label{eq:crps}
    \overline{CRPS}=\frac{1}{n}\sum_{i=1}^{n}
 \int_{-\infty}^{\infty}[F_i(x)-\mathbf{1}(x\geq y_i)]^2 dx
\end{equation}

When compared to a `reference' (see Section~\ref{data_methods}\ref{sec:benchmarking}) forecast, the continuous ranked probability skill score (CRPSS) can be derived as follows:

\begin{align*}
    CRPSS &= \frac{CRPS_{\text{reference}} - CRPS}{CRPS_{\text{reference}} - CRPS_{\text{perfect}}} \\
    &= 1 - \frac{CRPS}{CRPS_{\text{reference}}}
\end{align*}
where $CRPS\textsubscript{reference}$ is the CRPS of the reference forecast and $CRPS\textsubscript{perfect}$ is the CRPS of the perfect forecast, which is zero.

\subsubsection{Probability Integral Transform histogram}

A common method to evaluate the calibration of ensemble forecasts is through verification rank histograms \citep{anderson1996method, hamill1997verification}. These histograms assess the reliability of ensemble forecasts by comparing them with observations. If the ensemble forecast is well-calibrated, the resulting histogram should display a uniform distribution (flat) indicating that each bin represents an equally likely scenario. This concept can be extended to probabilistic forecasts by constructing Probability Integral Transform (PIT) histograms using a theorem that states that, when the cumulative distribution function of any random variable is treated as another random variable, the resultant distribution is uniform. Thus, PIT histograms for well-calibrated probabilistic forecasts will be flat, reflecting consistent density across all probabilities facilitating direct comparisons with verification rank histograms.

\subsection{Easy Uncertainty Quantification \label{easyUQ}}
The chaotic nature of the atmosphere motivates the use of probabilistic representations in weather prediction. Operational weather services currently address this need through the use of expensive NWP ensemble forecasts \citep{bauer2015quiet}. A novel tool, Easy Uncertainty Quantification \citep[EasyUQ; ][]{walz2024easy}, offers a cheaper alternative to this approach by transforming deterministic forecasts, such as those generated by ML or NWP models, into probabilistic forecasts based on past forecast performance. 

Notably, EasyUQ\footnote{The Python package to implement EasyUQ is available at \url{https://github.com/evwalz/easyuq.git}} only requires the deterministic forecast itself and corresponding observations without any detailed knowledge about the origin of the forecast. This feature makes EasyUQ a readily implementable and economically viable solution for generating probabilistic forecasts. EasyUQ employs Isotonic Distributional Regression  \citep[IDR; ][]{henzi2021isotonic} to generate discrete predictive distributions, eliminating the need for parameter tuning. An added advantage of IDR is that it provides readily calibrated outputs, thereby enhancing the reliability and usability of the probabilistic forecasts generated.

The steps involved in implementing EasyUQ are as follows: 1) Split the deterministic forecast into training-testing pairs, as explained in Section \ref{data_methods}\ref{pre_process}. 2) Train EasyUQ with the training data and generate probabilistic forecasts using the testing data. 3) As we operate in leave-one-out cross-validation, we repeat these processes for all training-testing splits to obtain probabilistic forecasts for the entire 2007-2019 period. We use CRPS as the metric for assessing the accuracy of the probabilistic forecast as recommended by \citep{walz2024easy}.

\subsection{Benchmarking \label{sec:benchmarking}}
To benchmark the probabilistic forecasts of rainfall accumulation based on ML models, we compare them with three others: 1) the Extended Probabilistic Climatology with a 15-day window \citep[EPC15;][]{walz2021imerg}, 2) the operational ECMWF ensemble daily precipitation forecast (ENS) and, 3) a probabilistic forecast generated by transforming the ENS control forecast using the aforementioned EasyUQ (CTRL EasyUQ).

The EPC15 is a climatology-based ensemble forecast that is computed using a 15-day window on either side of the target day, resulting in a total of 31 days. Viewed as an ensemble,  EPC15 comprises 372 members (31 days $\times$ 12 training years), which are calculated for all days from 2007 to 2019. In this study, we consider EPC15 as the main reference forecast for skill assessment. The ENS forecast is the 24-hour raw ensemble precipitation forecast from the ECMWF. The CTRL EasyUQ forecast is generated by transforming the ENS control member into a probabilistic forecast, as explained in Section~\ref{data_methods}\ref{easyUQ}. 

To determine whether the observed differences in the CRPSS are statistically significant, we use the Diebold-Mariano test \citep{diebold2002comparing} with a Benjamini-Hochberg correction \citep{benjamini1995controlling}, following the recommendation of \citet{wilks2011statistical}.


\section{Machine learning forecast models \label{ml_models}}
\subsection{Gamma-distributed Generalised Linear Model \label{gamma_regression}}

Ordinary linear regression provides a straightforward approach to forecasting by employing a linear combination of predictors. However, this assumes that changes in predictors induce proportionate linear changes in the target variable, which is inadequate for strictly non-negative quantities like rainfall accumulation, yielding suboptimal outcomes.

To address this, Generalized Linear Models \citep[GLMs;][]{nelder1972generalized, mccullagh1989generalized} use an inverse link function ($h$) to transform the linear combination of predictors (unknown parameters $\omega$ and predictors $X$) to appropriately relate the target variable ($\hat{y}$) to the predictors:

\begin{equation} \label{eq:link_func}
\hat{y}(\omega, X) = h(X\omega)
\end{equation}

Unlike ordinary linear regression, which uses the Identity function, GLMs offer the flexibility to select a suitable link function based on the target variable's characteristics. GLMs replace the conventional squared loss function with a customizable objective function, better adapting to the data:

\begin{equation} \label{eq:minimization}
\min_\omega\frac{1}{2n\textsubscript{samples}}\sum_i d(y_i,\hat{y}) + \frac{\alpha}{2}||\omega||^2_2,
\end{equation}
where $\alpha$ is a constant determining the strength of $L2$ regularization, $d$ represents the unit deviance, and $y_i$ denotes the observations.

Since rainfall accumulation is well approximated by a gamma distribution \citep{coe1982fitting, wilks1990maximum, sloughter2007probabilistic, scheuerer2015statistical}, a gamma-distributed GLM (gamma regression\footnote{Implemented using the \texttt{GammaRegressor} function from \texttt{scikit-learn version 1.2.2} \citep{pedregosa2011scikit}}) is suitable for forecasting rainfall amounts. The unit deviance $d$ in Equation \ref{eq:minimization} for gamma regression is defined as:

\begin{equation} \label{eq:unit_deviance}
d(y, \hat{y}) = 2\left(\log\frac{\hat{y}}{y} + \frac{y}{\hat{y}} - 1\right),
\end{equation}
following \citet{jorgensen1987exponential}. Since the gamma distribution fits only positive values, zero precipitation values are assigned a small positive quantity $\delta$ \citep{scheuerer2015statistical}:

\begin{equation} \label{eq:positive_observations}
y^{\prime} = \left\{
    \begin{array}{@{}rl@{}}
        y      & y > 0 \\
        \delta & y \leq 0, \\
    \end{array} 
\right.
\end{equation}

To forecast rainfall using gamma regression, the following steps are undertaken: 1) Fit the gamma regression model to the training dataset with $\alpha=0.4$, for a maximum of $10^6$ iterations or until minimum loss is attained within a tolerance of $10\textsuperscript{-4}$, using the L-BFGS-B algorithm \citep{zhu1997algorithm}. 2) Generate rainfall predictions by applying the trained model to the test dataset. Since we operate in leave-one-out cross-validation mode, these steps are repeated for every training-testing pair, and the results are concatenated for the $2007-2019$ period. 3) Apply EasyUQ (Section~\ref{data_methods}\ref{easyUQ}) to transform the deterministic forecast into a probabilistic forecast, referred to as the gamma regression forecast.

\subsection{Convolutional Neural Network Model \label{CNN}}
While the gamma regression model can capture simple non-linear relationships, it's efficacy diminishes when dealing with a large number of predictor variables. In contrast, neural networks offer a robust solution, adept at modelling intricate non-linear relationships across large predictor-target variable sets \citep{nielsen2015neural}. Specifically, convolutional neural networks (CNNs), tailored for structured, grid-like data, excel in this area. CNNs employ convolutions, which effectively mitigate overfitting by filtering input data through kernels to create feature maps, which efficiently connect the predictor and target data sets, enhancing computational efficiency and accuracy compared to fully connected neural networks.


In this study, we have chosen to employ a 1D CNN\footnote{Implemented using the \texttt{Conv1D} and associated functions from \texttt{keras v2.6.0 Python} package} architecture, driven by the utilization of zonally propagating TWs-based predictor variables. The architectural configuration of the 1D CNN model is rather simple. It is described as follows: Firstly, a 1D convolutional layer with 32 filters is implemented, employing a causally padded kernel size of length 3. Subsequently, a flattening layer is applied, succeeded by a hidden layer comprising 256 neurons and an output layer with a single neuron. To prevent overfitting, we employ a 25$\%$ Dropout \citep{srivastava2014dropout} in the hidden layer. All layers utilize the Rectified Linear Unit (\texttt{ReLu}) activation functions \citep{agarap2018deep} and adopt uniform kernel initialization. Other hyperparameters are left as default. 

\begin{figure}[!ht]
    \centering
    \noindent\includegraphics[width=0.85\textwidth]{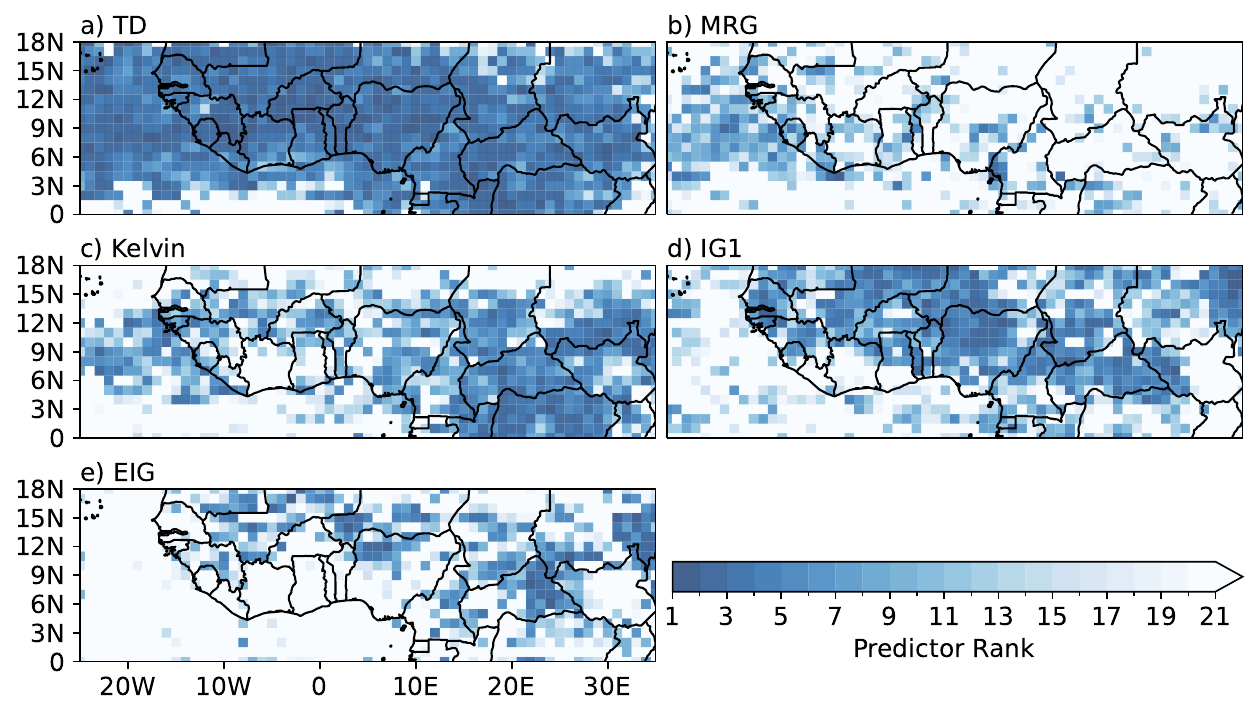}
    \caption{Ranks of PWAs for a) TD, b) MRG, c) Kelvin, d) IG1, and e) EIG waves (lower ranks denote higher relative importance) identified solely from 3 grid points downstream of the target grid point (D3) within the analysis domain during JAS from 2007 to 2019. We shade ranks only up to 21 out of 63 to highlight the most important predictors. We exclude MJO and ER due to their limited relevance for daily rainfall in tropical Africa.}
    \label{fig:wave_ranks_D3}
\end{figure}

The following steps are taken to produce CNN forecasts: 1) Scale predictor and target variables by dividing each by its maximum absolute value to ensure the highest value in the dataset is 1.0 without centering or shifting the data. 2) Partition the dataset into training and validation subsets, reserving the final year for validation. 3) Train the CNN model for a maximum of 500 epochs with a learning rate of $10^{-4}$, using mean absolute error (MAE) as the loss function. Implement early stopping based on the Taylor score (see Section~\ref{data_methods}\ref{metrics}) of the validation set, halting training if no improvement is observed for 5 consecutive epochs. 4) Use the \texttt{Adamax} optimizer, a variant of the \texttt{Adam} optimizer, to adaptively adjust the learning rate according to the data characteristics \citep{kingma2014adam}. 5) After training, generate predictions using the test data and re-scale the results to their original range. 6) Repeat these steps for every training-testing pair, concatenating the outcomes to obtain the full deterministic forecast for the 2007-2019 period (leave-one-out cross-validation). 7) Finally, apply EasyUQ to transform this deterministic forecast into a probabilistic forecast (referred to as CNN forecast unless stated otherwise)\footnote{The \texttt{Python} package to produce gamma regression and CNN forecasts is available at \url{https://github.com/athulrs177/forecast_models.git}}.

\section{Results \label{results}}

Our analysis begins with an examination of the spatial patterns of predictors identified by the predictor selection algorithm in subsection \ref{results}\ref{regional_predictor_selection}, which relies on gradient-boosting regression (see Section \ref{data_methods}\ref{grad_boost}). This offers insights into the most important predictors chosen to train the model across various regions within tropical Africa. The gamma regression and CNN models described in Section \ref{ml_models} were trained using these selected predictor variables. These models were then employed to generate forecasts of daily rainfall accumulation across tropical Africa for the period 2007-2019. In subsection \ref{results}\ref{results:forecast_niamey} we first evaluate the performance of these ML models in the vicinity of Niamey ($13^\circ$N, $2^\circ$E) to illustrate some key differences compared to the benchmark forecasts. Subsequently in subsection \ref{results}\ref{results:forecast_africa}, we extend our analysis to encompass the entire analysis domain.

\subsection{Relative predictor importance across tropical Africa \label{regional_predictor_selection}}

As outlined in Section~\ref{data_methods}, the selection algorithm described in Section~\ref{data_methods}\ref{grad_boost} identifies the optimal predictors for training the ML models based on their relative importance, as illustrated in Figure \ref{fig:feature_importance}. To visualize the spatial distribution of the importance of TW-based predictors across tropical Africa, we rank these predictors from 1 to 63 for each grid point, with lower ranks indicating higher importance. Figure \ref{fig:wave_ranks_D3} presents such an illustration exemplarily for predictor variables identified three grid points downstream from the target grid point (referred to as D3 in Fig.~\ref{fig:feature_importance}). We do not show the ranks of MJO and ER wave-based predictors, as they consistently attain very low ranks due to their limited impact on daily rainfall.

The most prominent predictor in D3 is TD across most of the analysis domain (Fig.~\ref{fig:wave_ranks_D3}a). This reflects the significant influence of AEWs on daily rainfall patterns in the Sahel \citep{fink2003spatiotemporal, schlueter2019systematic_b}. Similar patterns are observed at the target grid point, U9, and D5 (Figs.~\ref{fig:wave_ranks_target}a, \ref{fig:wave_ranks_U9}a, \ref{fig:wave_ranks_D5}a). However, the predictors at the target grid point and D5 show patchier signals, while U9 shows lower ranks with minimal signals over the Gulf of Guinea. The high-ranking TD signals over the Sahel can be attributed to AEWs, though TD signals further south remain ambiguous. Two possible explanations for this are: 1) Circulation anomalies related to AEWs extending into the southern hemisphere \citep{kiladis2006three} and, 2) Slow-cyclonic vortices described in \citep{knippertz2017meteorological}. The latter are also thought to be a cause for improved rainfall occurrence predictability over the Gulf of Guinea in \citet{rasheeda2023sources}.

MRG waves of slightly lower ranks are observed over the oceanic region off the coast of Guinea in D3 and D5 (Figs.~\ref{fig:wave_ranks_D3}b, \ref{fig:wave_ranks_D5}b) similar to the findings of \citet{schlueter2019systematic_b}. This region experiences very wet conditions in summer, with rainfall often?? triggered by the Guinea highlands. MRG waves, which have an off-equatorial convergence maximum, can project onto these rainfall signals, especially when other triggers are not as strong. Additionally, MRG waves can couple with AEWs to form MRG-AEW hybrids \citep{cheng2019two}, appearing as both MRG and AEW after filtering. However, MRG waves do not offer high predictability over land, as indicated by their low ranks.

Kelvin waves (Fig.~\ref{fig:wave_ranks_D3}c) show a less homogeneous pattern in D3 compared to TDs, with pronounced impacts over the eastern domain from the Congo Basin to the eastern Sahel. This broadly aligns with the findings of \citet{schlueter2019systematic_b}, but their study did not cover Central African regions south of $5^\circ$N. Higher ranks are also found over Nigeria, Benin, Burkina Faso, Mali, the West African coast, and the adjacent oceanic region. This aligns with prior studies \citep[e.g.,][]{sinclaire2015synoptic, mekonnen2016mechanisms}, which identify Kelvin waves as significant drivers of synoptic-scale rainfall in West and Central Africa. Unlike TDs, Kelvin wave influence ceases around $15^\circ$ N, consistent with its theoretical maximum at the equator. The coupling between AEWs and Kelvin waves, where one wave type can trigger the other, may explain the strong TD signal observed (Fig.~\ref{fig:wave_ranks_D3}a). Similar Kelvin wave patterns are also observed in D5 (Fig.~\ref{fig:wave_ranks_D5}c) but are patchier. 

IG1 waves show heightened activity throughout the Sahel and parts of Central Africa in D3 (Fig.~\ref{fig:wave_ranks_D3}d), despite theoretical expectations of a divergence maximum at the equator. \citet{jung2023link} used global simulations to filter IG1 waves with two methods: the FFT-based approach from \citet{wheeler1999convectively} for precipitation fields and the 2D pattern-projection method from \citet{yang2003convectively} for dynamical fields. Their methods effectively filtered IG1 waves only in precipitation data suggesting that the IG1 waves identified may not be true TWs but rather squall lines associated with MCSs, due to overlap in wavenumber and frequency. Similarly, the observed IG1 signals in this study may also be attributed to such phenomena, explaining why most high ranks are restricted to land regions. In comparison, D5 and U9 (Figs.~\ref{fig:wave_ranks_D5}d, \ref{fig:wave_ranks_U9}d) show similar but patchier patterns, with D5 showing comparable ranks and U9 lower ranks.

Similar to IG1, EIG waves in D3 (Fig.~\ref{fig:wave_ranks_D3}e) are mostly confined to land regions but show patchier patterns. D5 (Fig.~\ref{fig:wave_ranks_D5}e) and D7 (not shown) exhibit more robust patterns covering large portions of Central Africa and the western Sahel, while the target and U9 (Figs.~\ref{fig:wave_ranks_target}e and \ref{fig:wave_ranks_U9}e) show very low ranks. High ranks in EIG wave predictors align with regions of strong MCS activities. The eastward propagation of these signals is ambiguous, as MCSs propagate westward. One possible explanation is that EIG waves trigger MCSs, which then propagate westward. Doppler-shifted MRG waves \citep{yang2003convectively} is another potential cause.

\subsection{Forecast performance at Niamey \label{results:forecast_niamey}}

\begin{figure}[!ht]
    \centering
    \includegraphics[width=0.85\textwidth]{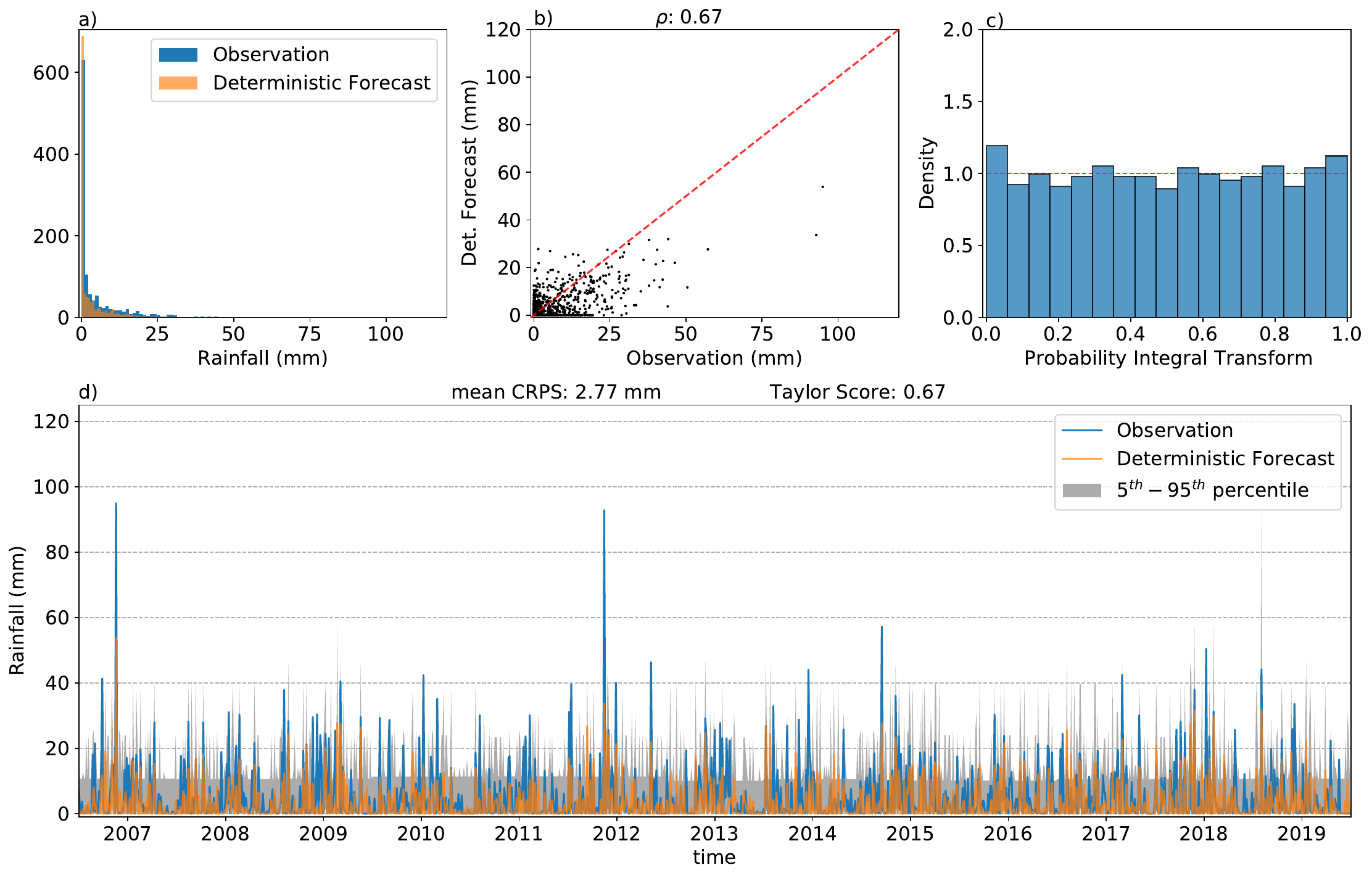}
    \noindent\caption{Illustration of CNN-based forecast of 24-hour rainfall accumulation at the grid point near Niamey ($13^\circ$ N, $2^\circ$ E) using TW-based predictors: a) A histogram depicting the distribution of observed (blue) and (deterministically) forecast (orange) rainfall; b) A scatter plot of observed (x-axis) and forecast (y-axis) rainfall with the Pearson correlation coefficient ($\rho$) between the two variables displayed in the title. The red dashed line along the diagonal denotes perfect correlation ($\rho=1$); c) A PIT histogram demonstrating the calibration of the CNN (probabilistic) forecast. The red dashed line represents a standard uniform distribution, serving as a reference for a perfectly calibrated forecast; d) A time series plot exhibiting observed (blue) and (deterministic) forecast (orange) rainfall for all JAS seasons from 2007 to 2019. The grey shading indicates the 5\textsuperscript{th} to 95\textsuperscript{th} percentile range of the CNN (probabilistic) forecast. Additionally, the mean CRPS of the probabilistic forecast and the Taylor Score of the deterministic forecast are also provided.}
    \label{fig:cnn_forecast_niamey}
\end{figure}

\begin{figure}[!ht]
    \centering
    \noindent\includegraphics[width=0.85\textwidth]{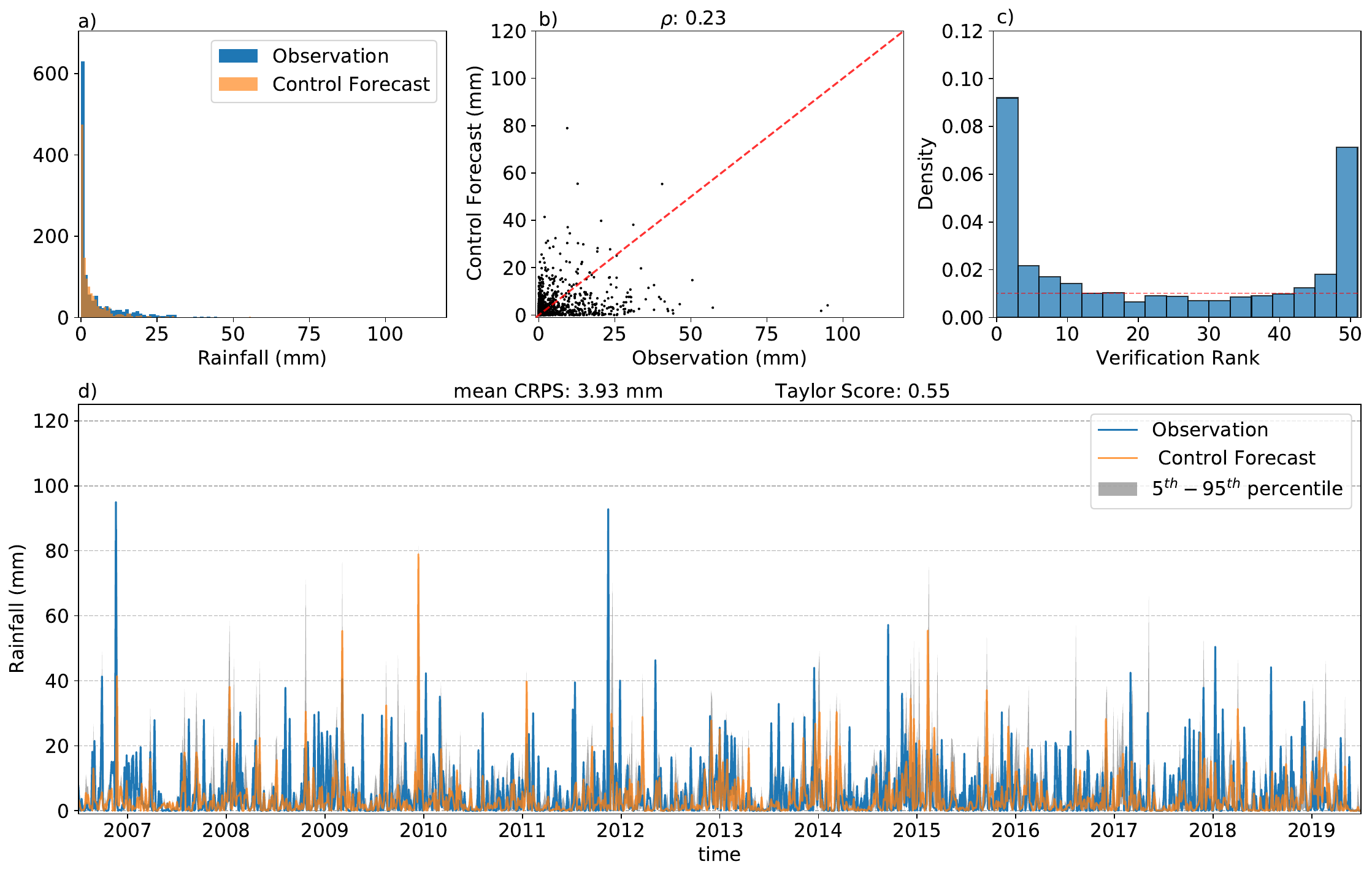}
    \caption{Same as Fig.~\ref{fig:cnn_forecast_niamey}, but for ENS forecast. Note that c) shows a verification rank histogram since ENS has a limited number (51) of ensemble members.}
    \label{fig:ens_forecast_niamey}
\end{figure}

Now, we illustrate the behavior and performance of the ML forecast models at the grid point near Niamey through a range of diagnostics as shown in Figs.~\ref{fig:cnn_forecast_niamey} and \ref{fig:ens_forecast_niamey}. Specifically, Figs.~\ref{fig:cnn_forecast_niamey}a and \ref{fig:ens_forecast_niamey}a show histograms of observed (blue) and deterministic forecast (orange) rainfall. The observed rainfall distribution has a predominance of very low to no rainfall occurrences (more than 400 out of 1196 timesteps), leading to a long right tail. Both the CNN model (more than 600 out of 1196 timesteps) and ENS (more than 200 out of 1196 timesteps) capture this characteristic with their right-tailed distributions. In contrast, the Gamma regression (zero no-rain days; see Fig.~\ref{fig:gamma_forecast_niamey}a) and EPC15 (zero no-rain days; see Fig.~\ref{fig:epc_forecast_niamey}a) fail to emulate this behavior. This is expected, as extreme values are lost in EPC15 computation, and Gamma regression does not predict zeros by design.

The scatter plot between the observed rainfall and the corresponding CNN deterministic forecast is presented in Figure \ref{fig:cnn_forecast_niamey}b. The red dashed line along the diagonal represents a perfect match. The CNN deterministic forecast demonstrates a relatively high correlation of $0.67$ with the observed rainfall, indicating strong predictive performance. However, the majority of scatter points lie below the diagonal, suggesting a tendency of the CNN model to underestimate rainfall, despite being the best in this study. In comparison, the ENS control forecast, depicted in Figure \ref{fig:ens_forecast_niamey}b, exhibits a much lower correlation of $0.23$, reflecting a greater mismatch between the predicted and observed daily rainfall patterns. The gamma regression deterministic forecast (Figure \ref{fig:gamma_forecast_niamey}b) performs better than the ENS control, with a correlation coefficient of 0.$43$, but falls short of the CNN model. Finally, the EPC15 model, shown in Figure \ref{fig:epc_forecast_niamey}b, has a correlation of $0.16$, highlighting its limited capability in accurately capturing the variability of observed daily rainfall. 

The PIT histogram of the probabilistic CNN forecast is shown in Fig.~\ref{fig:cnn_forecast_niamey}c. It exhibits a flat shape with only minor deviations attributable to sampling issues, indicating robust calibration. Conversely, the PIT histogram for the ENS forecast (Fig.~\ref{fig:ens_forecast_niamey}c) manifests a `U' shape, characteristic of substantial miscalibration (underdispersive). In comparison, the PIT histograms of gamma regression, CTRL EasyUQ, and the EPC15 (see Figs.~\ref{fig:gamma_forecast_niamey}c -- \ref{fig:epc_forecast_niamey}c) forecasts all portray a flat distribution, suggesting good calibration. This is expected as EPC15 is calibrated by construction, while both gamma regression and CTRL EasyUQ are calibrated due to the use of EasyUQ (see section~\ref{data_methods}\ref{easyUQ}). 

\begin{table}[t]
\caption{Mean CRPS of the probabilistic (and ensemble) and deterministic forecasts (middle column) and the Taylor Score for the deterministic forecasts only (right column). Note that for deterministic forecasts, the CRPS breaks down into the mean absolute error (MAE, shown in italics). The Taylor Score for EPC15 is computed using the ensemble mean. The best scores are in bold.}\label{tab:crps_taylor}
\begin{center}
\begin{tabular}{c c c}
\hline\hline
Forecast & CRPS (mm) & Taylor Score \\
\hline
CNN deterministic              & \textit{3.78} & \textbf{0.67} \\
Gamma regression deterministic & \textit{5.33} & 0.2           \\
ENS control                    & \textit{5.19} & 0.55          \\
EPC15                          & 3.61          & 0.05          \\
CNN probabilistic              & \textbf{2.77} & --            \\
Gamma regression probabilistic & 2.87          & --            \\
ENS                            & 3.93          & --            \\
CTRL EasyUQ                    & 3.49          & --            \\
\hline
\end{tabular}
\end{center}
\end{table}

Finally, the time series of daily observed rainfall (blue) and the corresponding CNN deterministic forecast (orange) are shown in Fig.~\ref{fig:cnn_forecast_niamey}d. The shaded grey area represents the 5\textsuperscript{th} to 95\textsuperscript{th} percentile range of the probabilistic forecast computed via EasyUQ. Additionally, the Taylor Score of the deterministic forecast (i.e., before applying EasyUQ)  and the mean CRPS of the probabilistic forecast are presented. Note that higher Taylor Scores imply better deterministic forecasts and lower CRPSs imply better probabilistic forecasts (see Section~\ref{data_methods}\ref{metrics}). The CRPS and Taylor scores of all forecasts are tabulated in Table~\ref{tab:crps_taylor} for easy reference. In terms of deterministic forecasts, CNN exhibits superior performance compared to others, with a Taylor score of 0.67, while the ENS forecast (see Fig.~\ref{fig:ens_forecast_niamey}d) exhibits slightly poorer performance reaching a score of 0.55. The gamma regression forecast (Fig.~\ref{fig:gamma_forecast_niamey}d) achieves only a score of 0.2 indicating a poor deterministic forecast. Possible reasons for this may be: 1) unrealistic values of rainfall values (more than 500 mm at times) and 2) median value of rainfall around 3 mm which is not representative of Niamey. Since EPC15 is an ensemble forecast we compute the Taylor score with the ensemble mean, which is understandably the worst (0.05) among the compared forecasts, as most of variations are lost in its computation. This trend is further observed when considering Mean Absolute Error (MAE), where the CNN deterministic forecast records the lowest value of 3.78 mm, followed by the ENS control forecast with a value of 5.19 mm and gamma regression forecast with a value of 5.33 mm. 

Nonetheless, probabilistic (and ensemble) forecasts illustrate an improvement over these metrics. Among the probabilistic forecasts, with a mean CRPS of 2.77 mm, the CNN forecast outperforms all others. The gamma regression forecast trails slightly behind the CNN forecast, boasting a mean CRPS of 2.87 mm. The ENS forecast emerges as the poorest performer, registering a mean CRPS of 3.93 mm. The CTRL EasyUQ forecast exhibits a marked improvement over the raw ENS forecast, with a mean CRPS of 3.49 mm. These suggest that a simple model like gamma regression that only learns the general trends of daily rainfall can be improved immensely when combined with EasyUQ but it is limited when combined with highly miscalibrated outputs from NWP models.

\subsection{Forecast performance in Tropical Africa\label{results:forecast_africa}}

\begin{figure}[ht]
    \centering
    \noindent\includegraphics[width=0.488\textwidth]{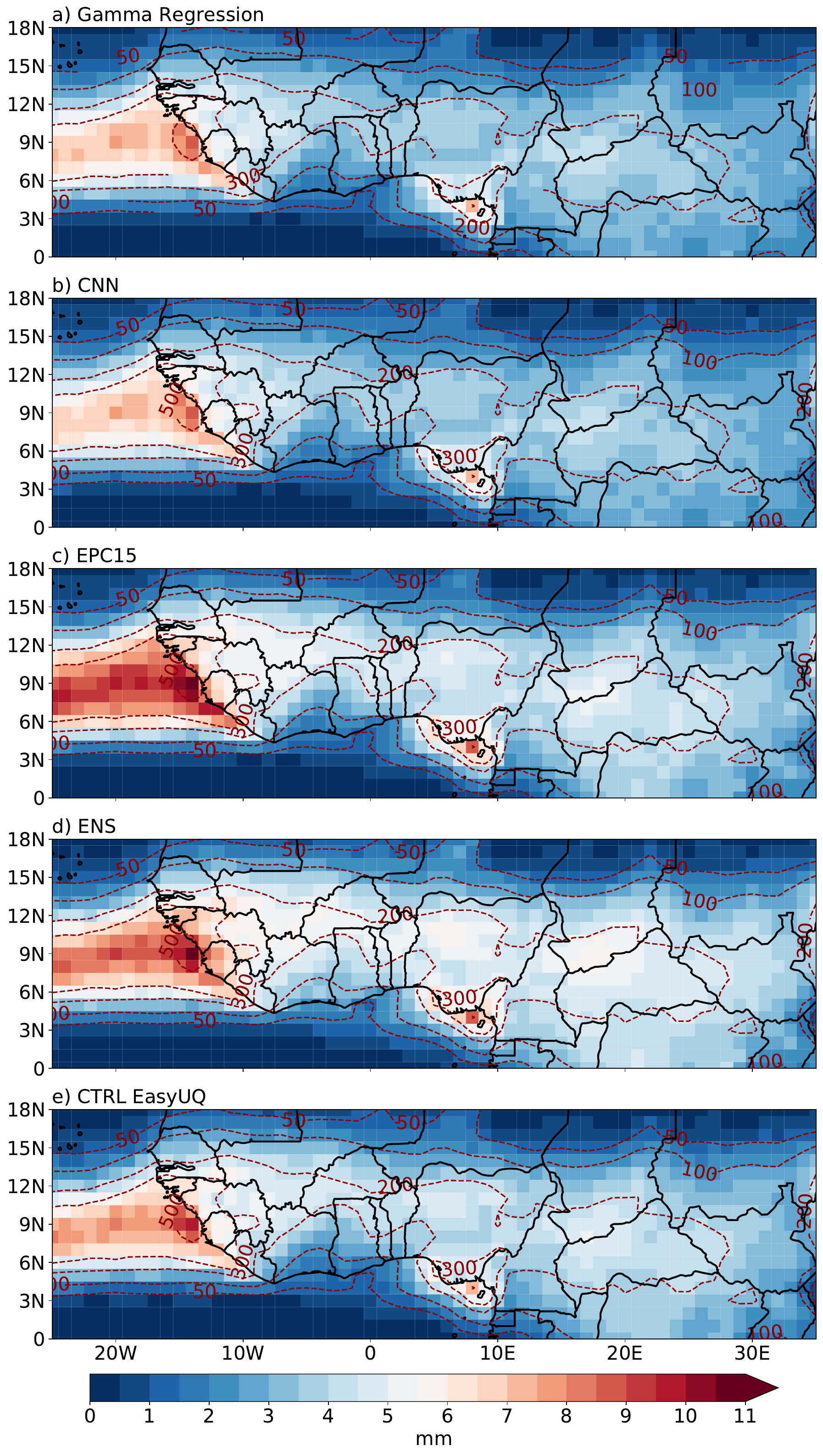}
    \caption{The CRPS (in mm) of a) Gamma regression, b) CNN, c) EPC15, d) ENS and e) CTRL EasyUQ forecasts for JAS from 2007~-2019. The contours indicate the mean monthly rainfall accumulation in mm.}
    \label{fig:crps_forecasts}
\end{figure}

We now move on to evaluate forecasts of daily rainfall using the ML models at all grid points in the nested analysis domain (see Fig.~\ref{fig:domain}) and compare their performances against the benchmark forecasts. To this end, firstly, we compute the CRPS of the forecasts across all grid points in the analysis domain as shown in Fig.~\ref{fig:crps_forecasts}.

The dark red contours in Fig.\ref{fig:crps_forecasts} illustrate the climatological rainfall pattern, expressed as monthly rainfall accumulation. These contours indicate that the coastal regions of Guinea and Cameroon receive the highest amounts of rainfall (exceeding 300 mm), likely orographically induced by the highland terrains in these regions (see Fig.\ref{fig:domain}), consistent with findings from previous studies \citep[e.g.,][]{nicholson2008intensity, nicholson2009factors}. Our analysis indicates that the CRPS (shading in Fig.~\ref{fig:crps_forecasts}) generally follows the rainfall pattern, as it is an absolute measure. This also indicates significant challenges in accurately predicting rainfall in these regions. Evidently, all models in our study struggle to produce accurate forecasts in the oceanic regions near Guinea. Conversely, the areas exhibiting the lowest CRPS values are situated over the Gulf of Guinea and regions to the north of the Sahel, with very little discernible disparities between different forecasts. 

The gamma regression forecast (Fig.\ref{fig:crps_forecasts}a) demonstrates relatively good performance over these regions, with CRPS values around 6-7 mm. Lower CRPS values are observed in other areas, mirroring the climatological rainfall patterns, with the lowest CRPS values (0-2 mm) found in the northeastern Sahel and equatorial Atlantic (monthly rainfall below 50 mm). Regions such as the western-central Sahel and the Congo Basin exhibit moderate CRPS values (2-5 mm). The CNN forecast (Fig.\ref{fig:crps_forecasts}b) shows a similar pattern to the gamma regression, suggesting that simpler statistical models can effectively capture most of the rainfall patterns associated with smooth predictors like TWs. However, the more sophisticated CNN model provides slightly better forecasts (lower CRPS by approximately 0.05 mm compared to gamma regression) in the African rainbelt region (including the Sahel, central Africa, and the oceanic region near Guinea), where rainfall is largely stochastic. Conversely, in regions with very low rainfall (e.g., northern Sahel and equatorial Atlantic Ocean), the skill of the CNN forecast is either worse or comparable to that of the gamma regression. 

The EPC15 forecast (Fig.~\ref{fig:crps_forecasts}c) demonstrates the poorest performance in the coastal region near Guinea, with CRPS values reaching up to 11 mm. This poor performance is particularly noteworthy because EPC15 is designed to be the most calibrated forecast in this study. The most likely reason for this poor skill is that, despite its calibration, the EPC15 forecast cannot capture sufficient variation since it is based on climatology. This results in very low discrimination capability, leading to poor CRPS. These findings are consistent with those of \citet{walz2024physics}, although their analyses were confined to land regions. Other regions exhibit similar patterns to the ML forecasts, following the climatological rainfall but with much higher CRPS values. For example, the northern region shows CRPS around 1 mm, the Sahel around 4.5-6 mm, and central Africa around 3-5.5 mm. The ENS forecast (Fig.\ref{fig:crps_forecasts}d) shows slightly better CRPS values (8-10 mm) in the coastal region near Guinea but worse CRPS values on land, particularly in the rainbelt regions (5.5-7 mm in the Sahel and around 5 mm in central Africa). This indicates the poor skill of the ENS forecast in the rainbelt, where rainfall is highly stochastic. Post-processing with EasyUQ improves the skill of ENS substantially over land, as evidenced by the improved CRPS values in ENS CTRL (around 3-4 mm in the Sahel and 2.5-3.5 mm in central Africa) as shown in Fig.\ref{fig:crps_forecasts}e however, these are only small improvements over the EPC15 forecast.



\begin{figure}[ht]
    \centering
    \noindent\includegraphics[width=0.488\textwidth]{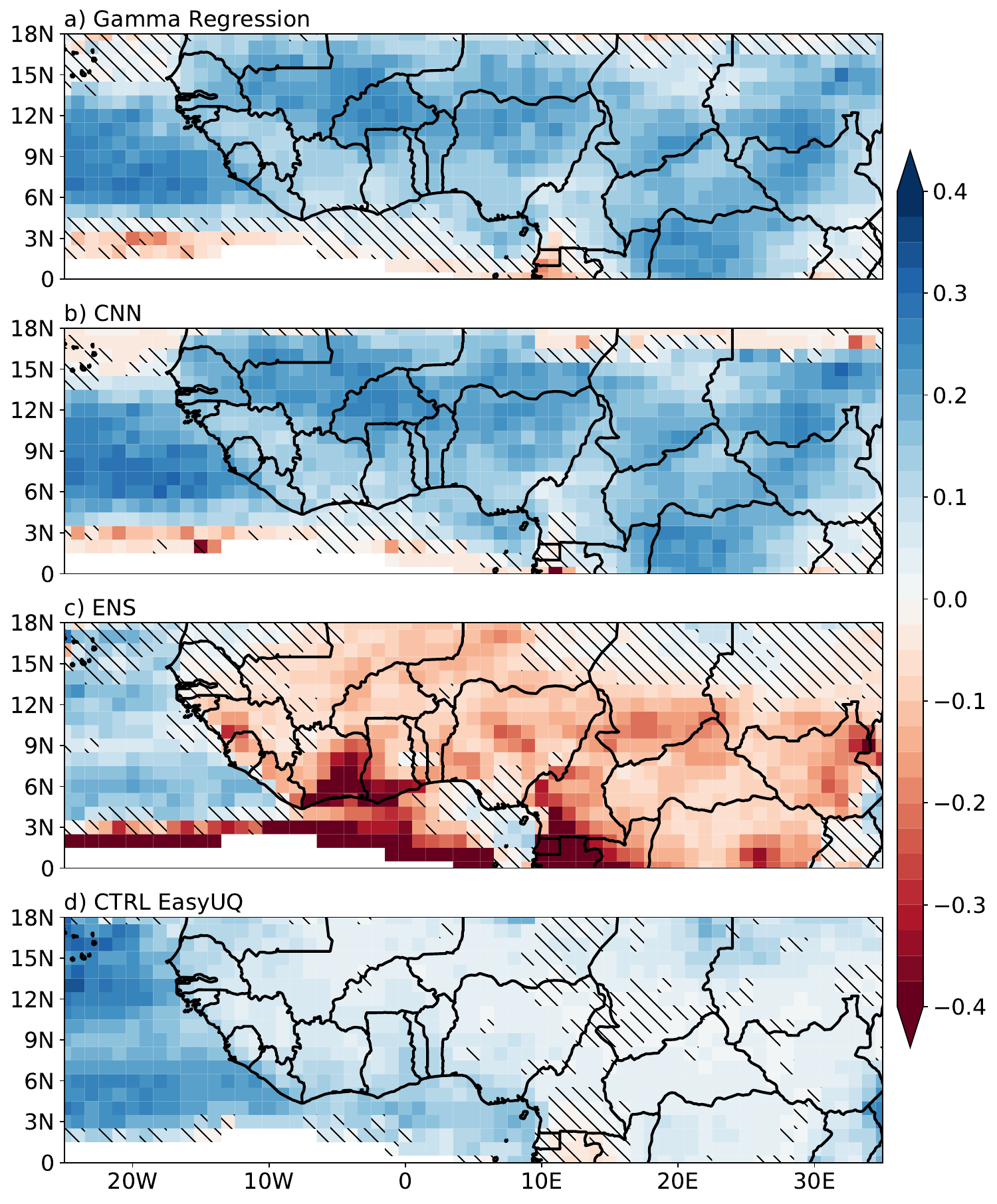}
    \caption{The CRPSS of a) Gamma regression, b) CNN, c) ENS, and d) CTRL EasyUQ forecasts relative to the EPC15 forecast for the JAS period spanning from 2007 to 2019. Hatching indicates grid points where the difference in the mean CRPS fails to be statistically significant at a significance level of $5\%$, determined using the Benjamini-Hochberg corrected Diebold-Mariano test for equal predictive ability. Grid points with mean monthly rainfall less than 5 mm are left blank to highlight regions of higher forecasting relevance.}
    \label{fig:crpss_forecasts}
\end{figure}

To bring out the skill improvement of the new ML forecast model more clearly, we calculate the CRPSS of all forecasts using EPC15 as the reference, as shown in Fig.~\ref{fig:crpss_forecasts}. This approach also facilitates direct comparison with the findings of \citet{walz2024physics}.

The CRPSS of the gamma regression forecast (Fig.\ref{fig:crpss_forecasts}a) exhibits substantial improvement over the EPC15 forecast across the majority of the analysis domain (blue shading). The largest improvements are observed over the oceanic region near Guinea, the western Sahel, and the Congo Basin. Regions in the far northeast and the coastal areas show the lowest improvements. Some parts of the analysis domain also demonstrate poorer performance compared to the EPC15 forecast: the oceanic region near the coast of Senegal, some grid boxes in the far northeast, and a large region in the Gulf of Guinea. However, the drop in skill is not statistically significant in these regions according to the Diebold-Mariano test \citep{diebold2002comparing} for equal predictive ability  (see hatching in Fig.\ref{fig:crpss_forecasts}a). That said, a likely reason for the reduced skill may be the very low rainfall in these regions, leading to suboptimal filtering of TWs. The CRPSS of the CNN forecast (Fig.~\ref{fig:crpss_forecasts}b) also demonstrates very similar patterns but shows slight improvement over the gamma regression, especially over the western and eastern Sahel, and the Congo Basin. The improvement over the EPC15 forecast in the oceanic region near Guinea also extends slightly more equatorward. However, the CNN forecast performs worse in regions of very low rainfall, such as the coast of Senegal, the Gulf of Guinea, and the far northeastern Sahel. In regions where TWs are strong and prevalent, the CNN possibly learns sophisticated patterns and may offer greater skill. However, in regions where the signals are weak and noisy, the CNN model fails to generalize and thus provide worse predictions compared to a simpler model.

The results obtained from the CNN forecast in our investigation (Fig.~\ref{fig:crpss_forecasts}b) are directly comparable to the CNN forecast depicted in Figure 11f of \citet{walz2024physics}, despite employing different model architectures and cross-validation strategies. While the CNN in \citet{walz2024physics} demonstrates superior performance along coastal regions, our CNN model showcases enhanced performance further inland, particularly over the Sahel and central Africa. The former may be a reason of utilizing a 2D CNN by \citet{walz2024physics}, which captures features dominated by topography; e.g., along the Guinea Coast where TW predictors (except TD) do not contribute much to the predictability as shown in Section~\ref{results}\ref{regional_predictor_selection}. This also explains the lower skill in our models in this region. We attribute the improved skill further inland to the inclusion of explicit zonally propagating TW information during the training of our 1D CNN models. Such information may not be as discernible in the version employed by \citet{walz2024physics} as their model was trained solely on previous days' rainfall and other meteorological variables. 

The CRPSS of ENS (Fig.~\ref{fig:crpss_forecasts}c) highlights the poor skill (red shading) of ensemble precipitation forecasts over tropical Africa, aligning with the conclusions drawn in several previous studies \citep[e.g., ][]{vogel2020skill, vogel2021statistical, rasheeda2023sources}. The EPC15 outperforms the ENS forecast across almost all land regions, particularly along the Guinea Coast and the coastal regions of Cameroon and Gabon, where CRPSS values are as low as -0.4. Over the Sahel and central Africa, where rainfall is largely stochastic, the ENS forecast exhibits slightly better skill but still underperforms compared to EPC15, with CRPSS values around -0.1 to -0.2. The ENS forecast shows positive skill (blue shading) over the oceanic region west of the mainland and parts of the northern Sahel, but these improvements are not statistically significant (hatching in Fig.~\ref{fig:crpss_forecasts}c). 

Although the postprocessing using EasyUQ significantly enhances the skill of ENS (Fig.~\ref{fig:crpss_forecasts}d), the improvement is only marginal over EPC15 indicated by the light blue shading over the land. Moreover, in several regions, such as central to eastern Sahel, southern Cameroon, and Gabon, the improvement lacks statistical significance (see hatching in Fig.~\ref{fig:crpss_forecasts}d). However, the improvement over the oceanic region seems to be substantial. By comparing the CRPSS of ENS and CTRL EasyUQ to their counterparts in \citet{walz2024physics} (see their Figs. 11a and 11c, respectively), we notice remarkable similarities. This reinforces the arguments for statistically postprocessing NWP precipitation forecasts in tropical Africa as it can already improve the calibration significantly.   

\section{Conclusions \label{conclusions}}

Current NWP models demonstrate limited skill in providing daily weather forecasts for tropical Africa, in particular with respect to rainfall \citep{vogel2018skill}. Studies employing simple statistical models to more advanced ML methods have proven to provide rainfall forecasts exceeding the skill of current ensemble prediction systems in tropical Africa \citep{vogel2021statistical, rasheeda2023sources, walz2024physics, antonio2024postprocessing}. However, no research has yet attempted to achieve the same by exploiting the predictive potential contained in TWs \citep[e.g.,][]{sinclaire2015synoptic, mekonnen2016mechanisms, schlueter2019systematic_b, schlueter2019systematic_a, ayesiga2021observed} within an operational context.
To explore this potential, we use satellite-based gridded precipitation product GPM IMERG version-7 \citep{huffman2023imerg} remapped to $1^\circ$ spatial resolution. First, we generate predictor variables that highlight the local impact of TWs \citep{van2016modulation, janiga2018subseasonal}, at a given target grid point. We also encompass predictors from grid points situated 3, 5, 7, and 9 grid points upstream and downstream, aligning with the direction of TW propagation. Subsequently, we train two ML models-- a simple gamma regression model and a more sophisticated CNN model, utilizing these predictor variables to issue predictions of daily rainfall accumulation with a 6-hour lead time in tropical Africa. Additionally, to maximise the accuracy of the ML model-based forecasts, we implement a predictor selection algorithm using gradient boosting regression to identify the most suitable set of predictors.

Most of the predictability stems from predictor variables computed at the downstream grid points. As we move further away from the target grid point, the importance of downstream predictors decreases, while that of upstream predictors increases, although not to the extent of the downstream grid points. TD-based predictors, particularly from proximal downstream and far away upstream grid points, emerge as the most crucial predictors across tropical Africa. While this evidence strongly suggests the significant modulation of rainfall by AEWs in the Sahel \citep{fink2003spatiotemporal}, the reason behind their importance in other regions may be attributed to southward extending AEWs and/or slow cyclonic vortices identified by \citet{knippertz2017meteorological}. 
MRG wave-based predictors are less prominent in comparison to TDs but exhibit reasonable signals downstream, particularly in the coastal region near Guinea. Since MRG waves have an off-equatorial convergence maximum, their signals projecting onto highland terrain-induced precipitation can be one reason for this pattern. Another potential reason may be the presence of AEW-MRG hybrid waves as described by \citet{cheng2019two}. However, only weak MRG signals are observed over land.

Kelvin wave-based predictors, particularly downstream, dominate the Central African region and parts of the eastern Sahel although less homogeneous compared to TDs broadly consistent with earlier studies \citep[e.g., ][]{sinclaire2015synoptic, mekonnen2016mechanisms}. The influence of Kelvin waves, ceases around $15^\circ N$, consistent with their theoretical maximum at the equator. A coupling between TDs and Kelvin waves \citep{lawton2022influence} resulting in the triggering of Kelvin waves due to the passage of a TD and vice versa may be a potential reason behind the overwhelming TD and Kelvin wave signals in Central Africa. Predictors based on IG1 waves are prominent in the Western-Central Sahel despite a theoretical divergence maximum at the equator, consistent with studies like \citet{jung2023link}, suggesting a coupling with MCSs. No discernible signals are observed over the ocean. 

Patchy signals of EIG waves can be observed in the Sahel and Central Africa in proximal downstream predictors with more robust patterns observed from further downstream attributed with high MCS activities observed over these regions. While MCSs propagate westward, the eastward propagating EIG waves may act as trigger for their initiation.  MJO and ER wave-based predictors are either not selected during predictor selection or offer minimal contribution. We posit that the predominance of downstream predictor variables is due to the negative manner in which predictability is achieved. TWs propagating away from the target grid point carry moisture along with the instability and convergence away, thereby reducing the potential for rainfall at the target grid point. While TWs can bring moisture from upstream grid points, we argue that this mechanism does not offer as high predictability.

We then compared the performance of the ML models with three benchmark forecasts: the climatology-based EPC15, the NWP-based ENS and the probabilistic CTRL EasyUQ. All models struggle in the oceanic regions near the coast of Guinea and Cameroon, where precipitation is largely influenced by proximity to highland terrain. Nevertheless, compared to the EPC15 forecast as reference, both ML models demonstrate clear performance gains across almost the entire analysis domain and outperform the benchmarks: the CNN forecast performs the best in the Sahel and parts of the Congo Basin, while the Gamma regression forecast performs better in arid regions in the North. The CRPSS of the ENS forecast underscores the limited performance of NWP forecasts in tropical Africa, consistent with findings from previous studies \citep[e.g.,][]{vogel2018skill, walz2021imerg, rasheeda2023sources}. The performance of raw ensemble precipitation forecast is sub-par in tropical Africa due to high miscalibration, consistent with the results of \citet{vogel2018skill, vogel2021statistical} and \citet{rasheeda2023sources}. Tools like EasyUQ can be effectively utilized to generate calibrated probabilistic forecasts from NWP control forecasts, offering a cost-effective solution, however only with marginal improvement to EPC15 when considering the cost of issuing ENS. While employing EasyUQ enhances the performance of the deterministic control (CTRL) forecast considerably, the gain is only marginal and fails to achieve statistical significance in several parts of the Sahel and Central Africa. Therefore, our primary recommendation is to leverage ML models in conjunction with tools like EasyUQ to produce probabilistic forecasts exceeding the skill of even postprocessed NWP-based forecasts. If developing a CNN model is unfeasible, a simpler gamma regression model suffices, as we have demonstrated that it outperforms the benchmarks and offers forecasts almost as skilful as the more sophisticated CNN.

The emergence of AI weather prediction models promises significant advancements in operational weather forecasting.  Our study provides a solid foundation for integrating precipitation forecasts into such systems. Leveraging advanced learning algorithms and extensive data, these models demonstrate impressive capabilities in predicting various meteorological variables. The recently introduced ECMWF AIFS, offers global forecasts of total precipitation, along with other variables with lead times up to 10 days improving upon the deficiencies of the earlier attempts like FourCastNet \citep{pathak2022fourcastnet}, Pangu-Weather \citep{bi2023accurate} and GraphCast \citep{lam2023learning}. However, their skill in challenging regions like tropical Africa remains to be investigated. By offering a comprehensive understanding of the advantages and limitations of purely data-driven precipitation forecasting techniques, we contribute to ongoing efforts to enhance these models. In particular, for tropical Africa, where rainfall is influenced by both (supposedly unpredictable) stochastic convection and (supposedly predictable) linear wave modes, knowledge of the latter, coupled with adequate (even simple) training approaches, can be highly effective. The main advantages of these approaches are: a) they are more cost-effective than running an NWP model with multiple ensemble members, and b) since the predictors are large-scale features and well-understood, the results can be interpreted physically. Although not strictly investigated within an operational framework, the gamma regression and CNN models introduced in this study can be trained and deployed operationally with minimal modifications by African weather services. These models can be further developed by incorporating weather station data to overcome the limitations of IMERG. Further studies are needed to extend this approach to longer lead times. Ultimately, our objective is to pave the way for the seamless integration of precipitation forecasts into AI weather models, thereby enhancing overall forecasting capabilities and the robustness of weather prediction systems. 

\clearpage
\acknowledgments

The research leading to these results has been accomplished within phase 2 of project C2, “Statistical-dynamical forecasts of tropical rainfall” of the Transregional Collaborative Research Center SFB/TRR165 “Waves to Weather” funded by the German Science Foundation (DFG). The authors thank their colleagues Tilmann Gneiting, Eva-Maria Walz and Benedikt Schulz for valuable discussions. 

An earlier version of the results presented are part of Athul Rasheeda Satheesh's PhD thesis.

%
%
\datastatement
The GPM-IMERG v7 data was obtained from NASA Global Precipitation Measurement Mission, available online at \url{https://disc.gsfc.nasa.gov/datasets/GPM_3IMERGHH_07/summary?keywords=IMERG}. The 24-hour precipitation ensemble forecast data was obtained from the TIGGE archive available online at \url{https://apps.ecmwf.int/datasets/data/tigge/levtype=sfc/type=pf/}. TIGGE (The Interactive Grand Global Ensemble) is an initiative of the World Weather Research Programme (WWRP).   


\bibliographystyle{ametsocV6}
\bibliography{references}

%

\newpage
\appendix




\begin{figure}[ht!]
    \centering
    \noindent\includegraphics[width=0.48\textwidth]{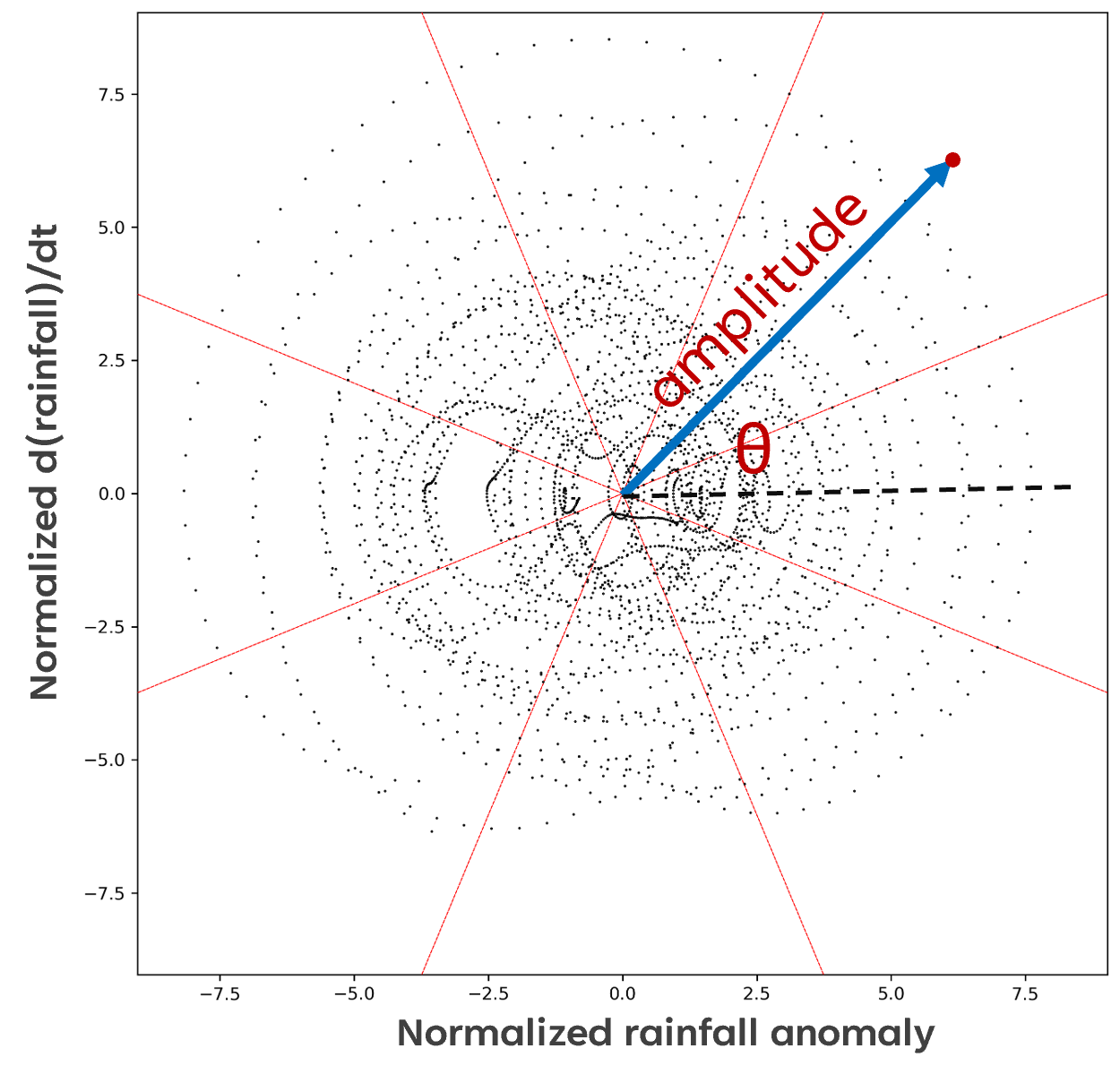}
    \caption{A schematic illustrating the computation of phase and amplitude of a wave mode. The x and y-axes show normalized rainfall anomaly and the time derivative of rainfall anomaly, respectively. Each scatter point represents a timestep.}
    \label{fig:phase_amplitude}
\end{figure}

\begin{figure*}[t!]
    \centering
    \noindent\includegraphics[width=0.75\textwidth]{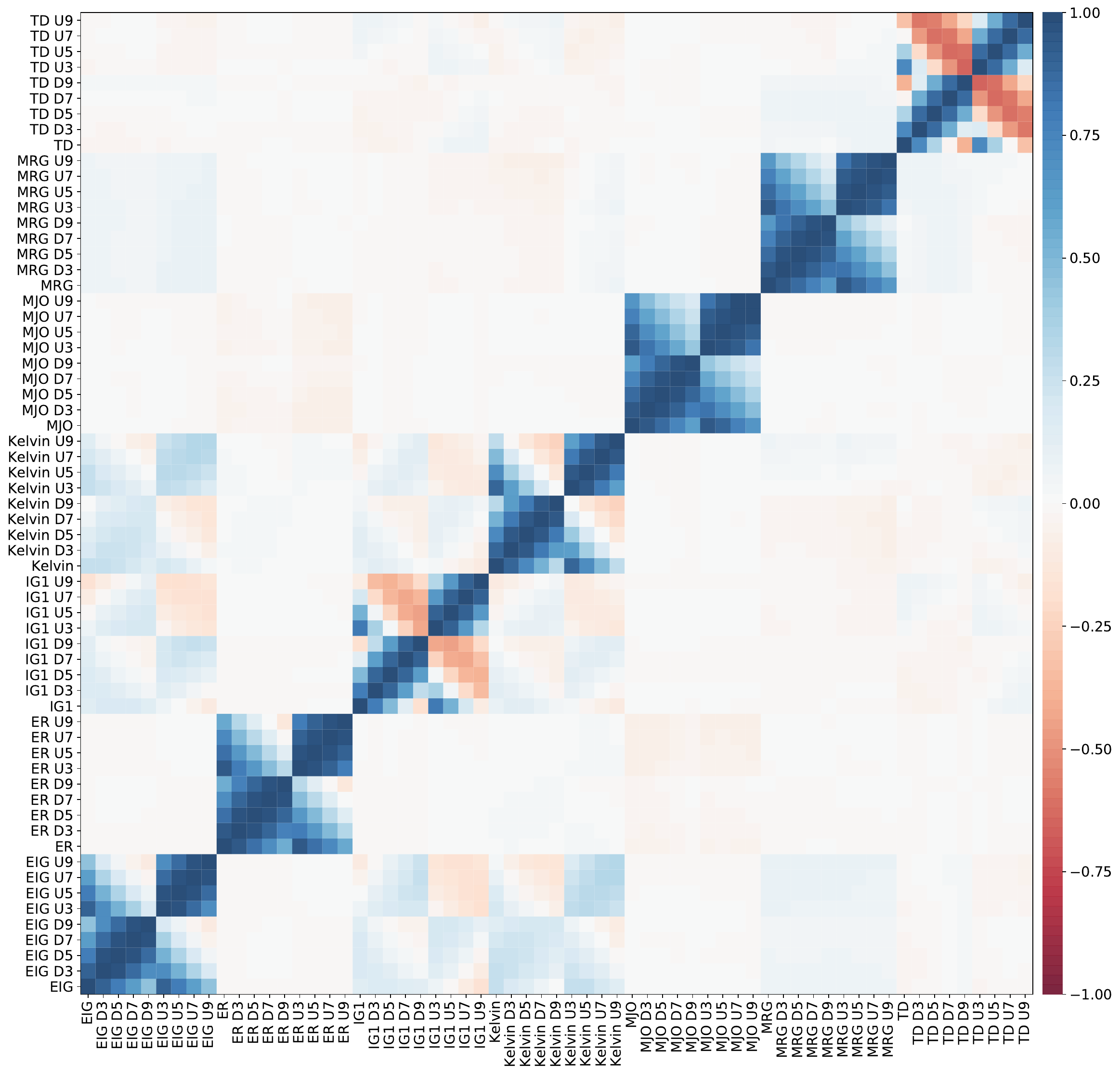}
    \caption{Cross-correlation (Spearman) map of PWAs at the grid point near Niamey ($13^\circ$N, $2^\circ$E). Blue (red) shading indicates positive (negative) correlation.}
    \label{fig:feature_correlation}
\end{figure*}

\begin{figure*}[!ht]
    \centering
    \noindent\includegraphics[width=0.75\textwidth]{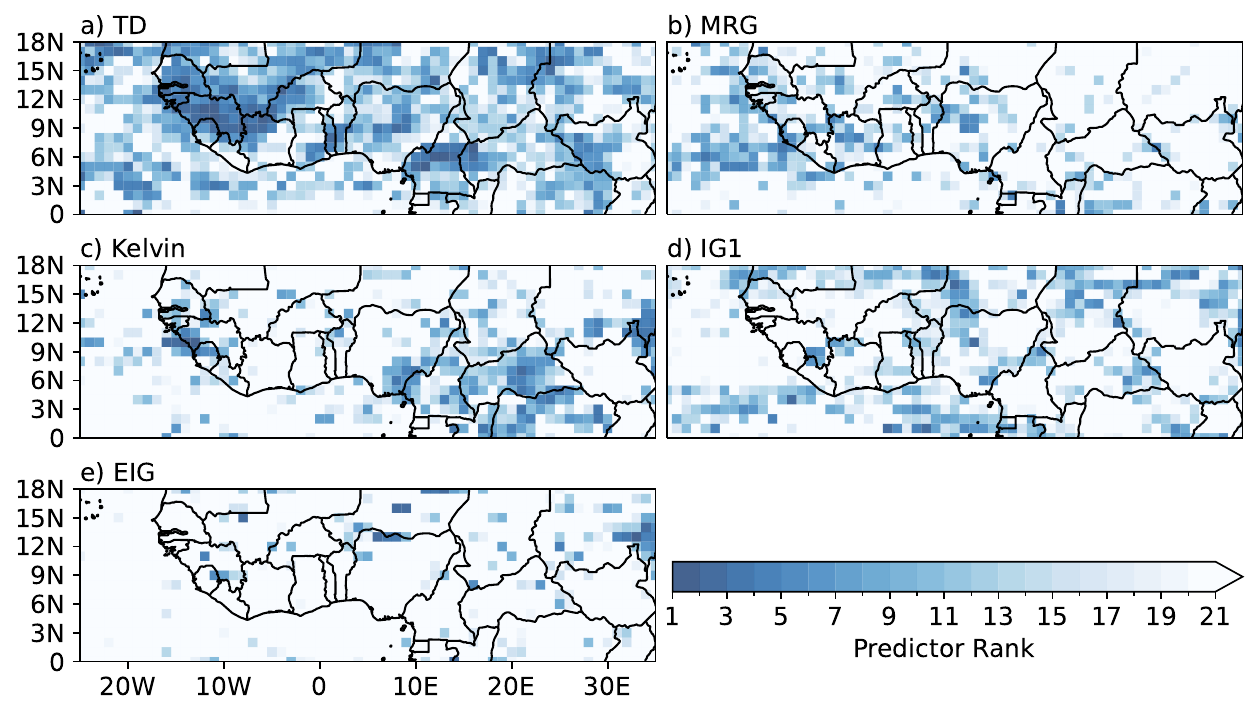}
    \caption{Same as Fig.~\ref{fig:wave_ranks_D3} but for target grid point.}
    \label{fig:wave_ranks_target}
\end{figure*}

\begin{figure*}[!ht]
    \centering
    \noindent\includegraphics[width=0.75\textwidth]{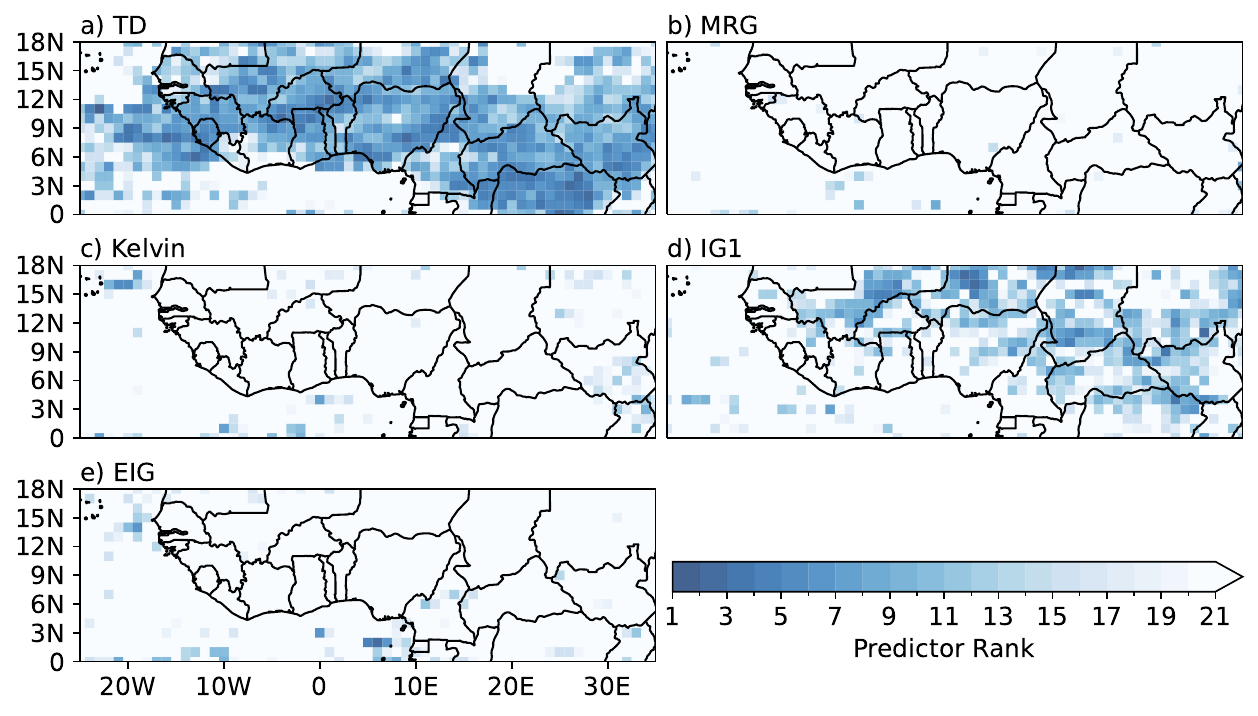}
    \caption{Same as Fig.~\ref{fig:wave_ranks_D3} but for U9.}
    \label{fig:wave_ranks_U9}
\end{figure*}

\begin{figure*}[!ht]
    \centering
    \noindent\includegraphics[width=0.85\textwidth]{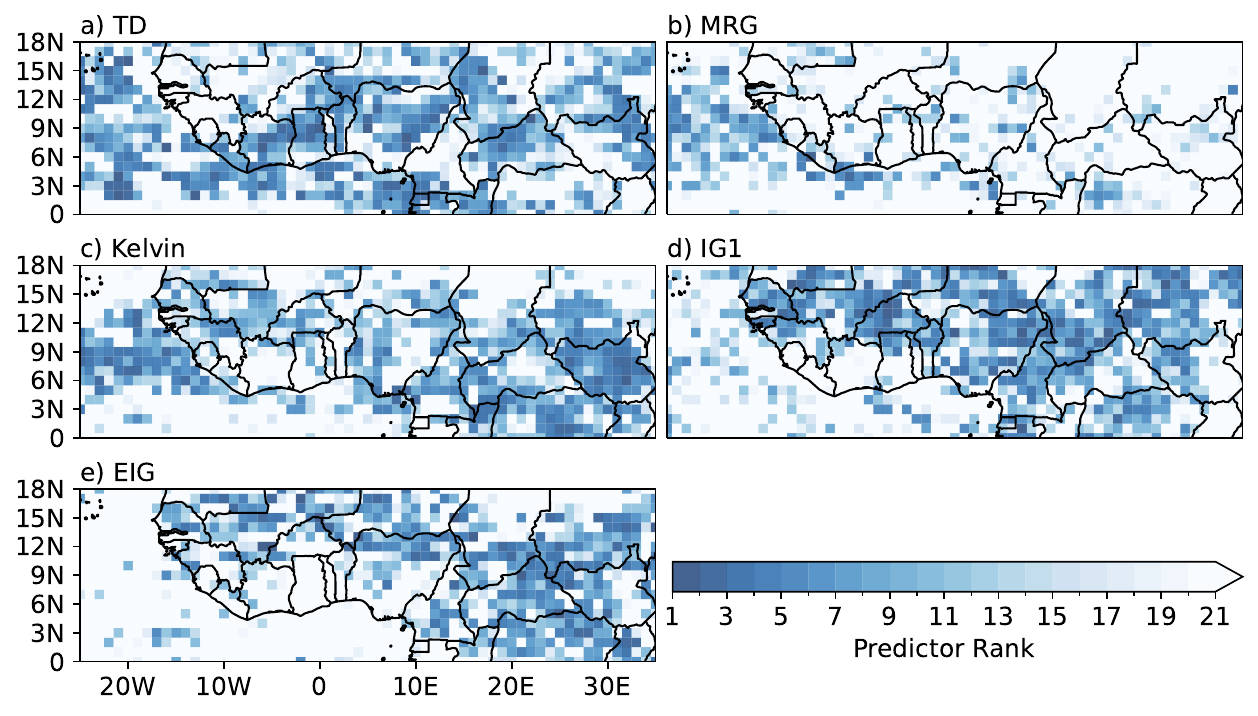}
    \caption{Same as Fig.~\ref{fig:wave_ranks_D3} but for D5.}
    \label{fig:wave_ranks_D5}
\end{figure*}

\begin{figure*}[!ht]
    \centering
    \noindent\includegraphics[width=0.75\textwidth]{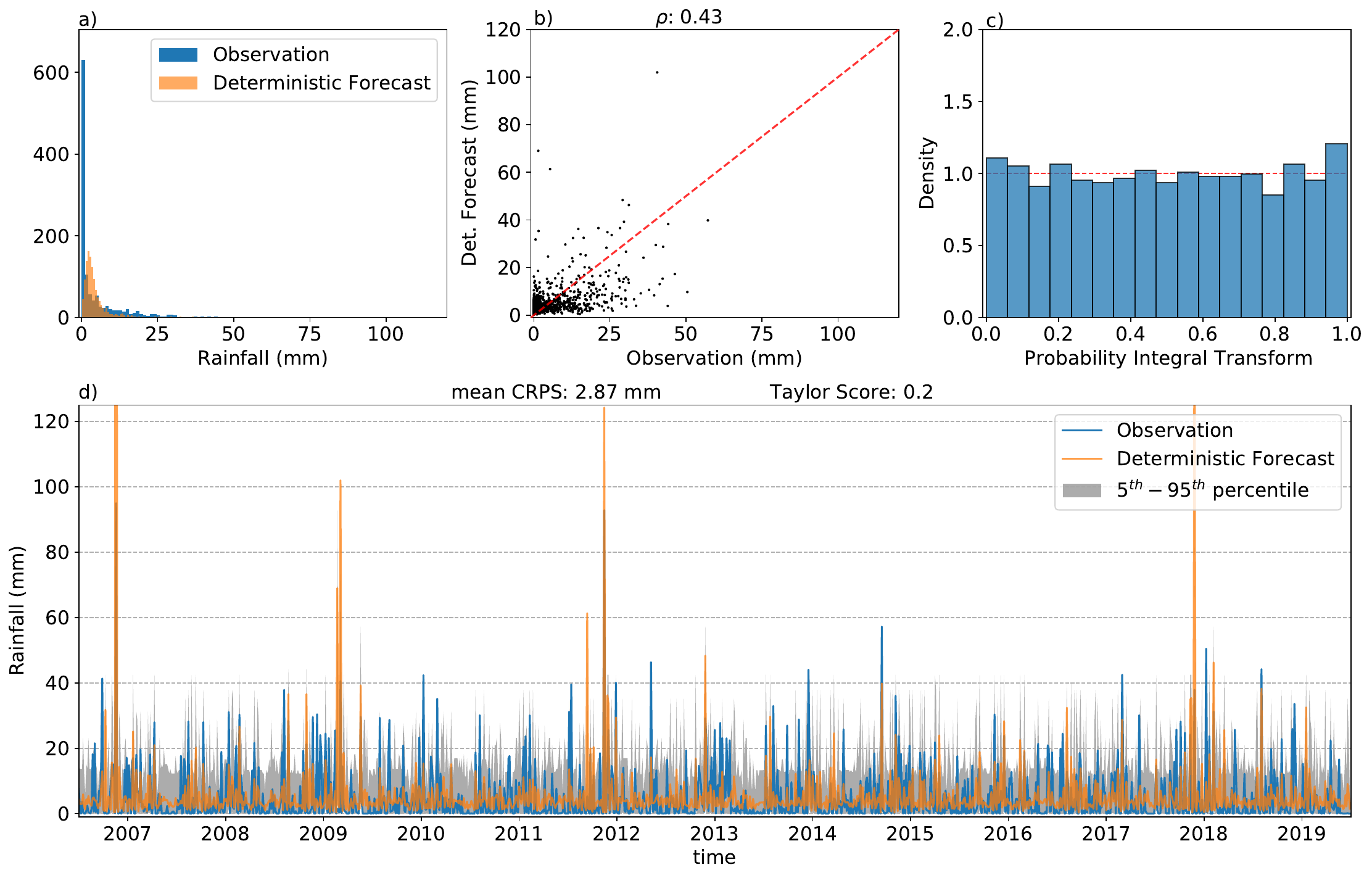}
    \caption{Same as Fig.~\ref{fig:cnn_forecast_niamey}, but for Gamma regression forecast.}
    \label{fig:gamma_forecast_niamey}
\end{figure*}

\begin{figure*}[!ht]
    \centering
    \noindent\includegraphics[width=0.75\textwidth]{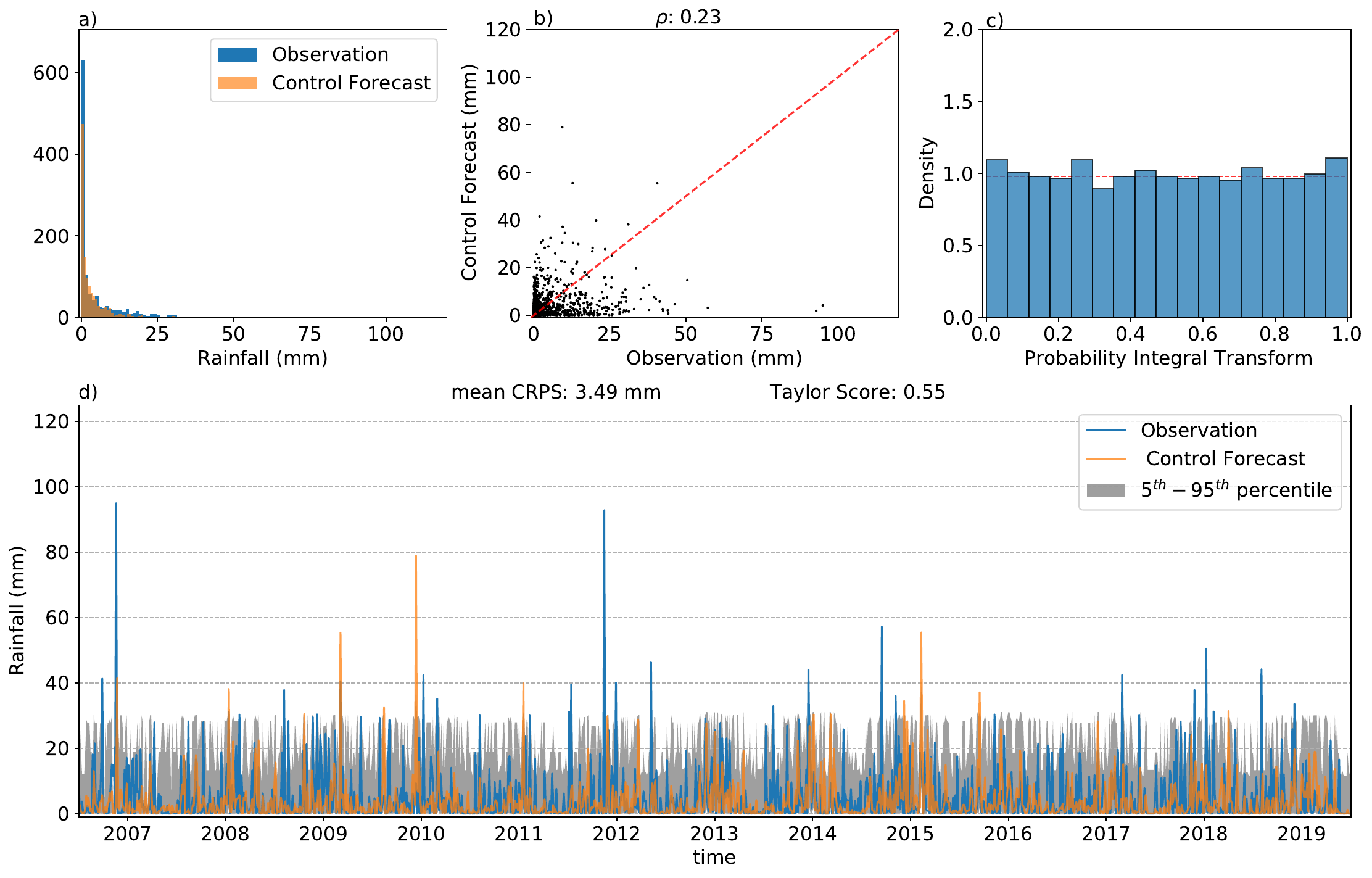}
    \caption{Same as Fig.~\ref{fig:cnn_forecast_niamey}, but for CTRL EasyUQ forecast.}
    \label{fig:ens_easyUQ_forecast_niamey}
\end{figure*}

\begin{figure*}[!ht]
    \centering
    \noindent\includegraphics[width=0.75\textwidth]{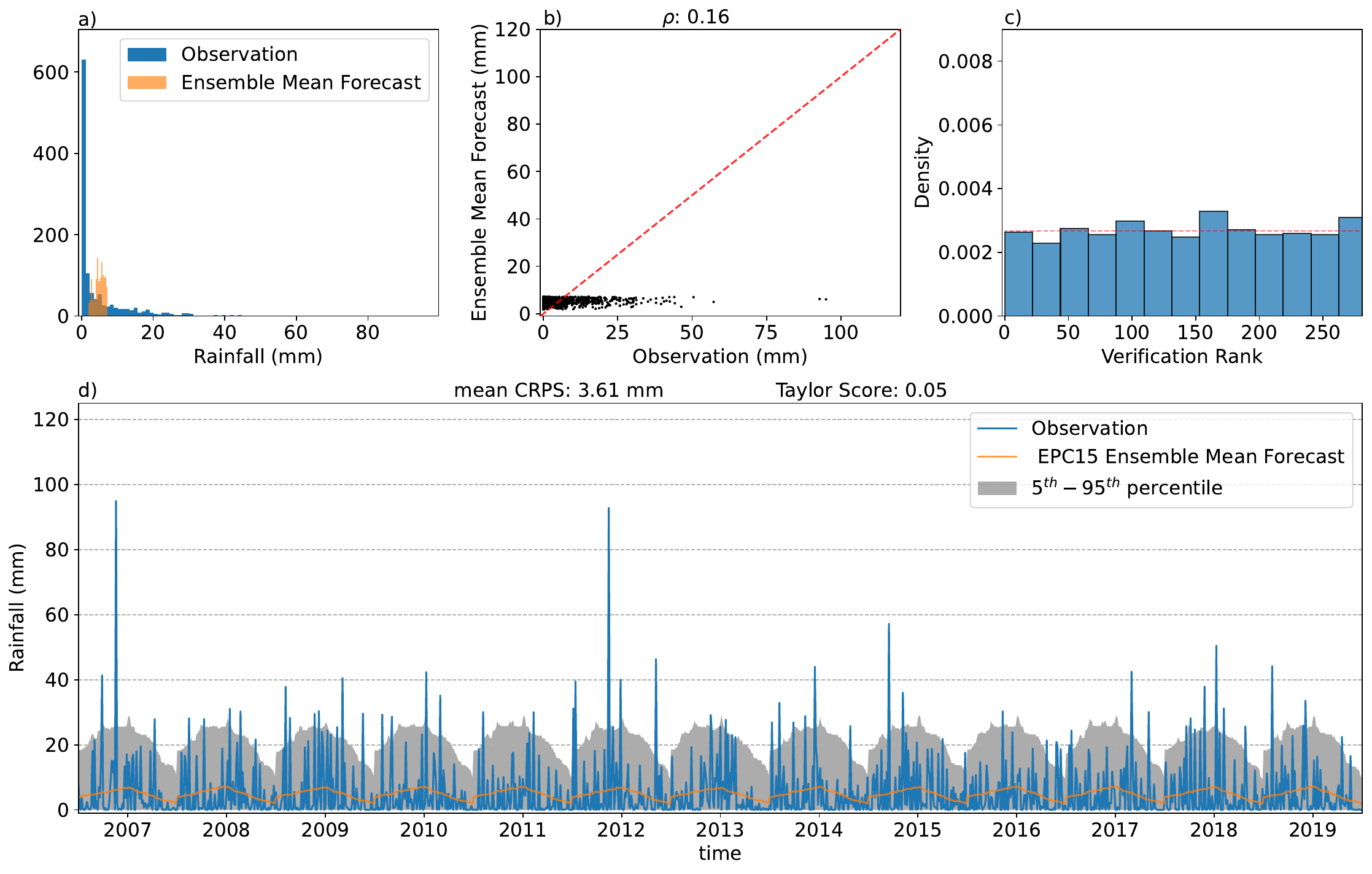}
    \caption{Same as Fig.~\ref{fig:cnn_forecast_niamey}, but for EPC15 forecast.}
    \label{fig:epc_forecast_niamey}
\end{figure*}
%



\end{document}